\documentstyle[12pt, preprint, aps, floats, epsfig, amsmath, amsbsy]{revtex}
\tighten

\newcommand{\nn}{\nonumber}

\newcommand{\rb}[1]{\raisebox{1.1ex}[-1.1ex]{#1}}
\newcommand{\lb}[1]{\raisebox{-1.1ex}[1.1ex]{#1}}
\newcommand{\bea}{\begin{eqnarray}}
\newcommand{\eea}{\end{eqnarray}}
\newcommand{\beq}{\begin{equation}}
\newcommand{\eeq}{\end{equation}}

\newcommand{\s}{\hat{s}}

\newcommand{\Order}{{\mathcal O}}
\newcommand{\bra}{\langle}
\newcommand{\ket}{\rangle}

\newcommand{\wtC}{\widetilde{C}}
\newcommand{\wtO}{\widetilde{O}}
\newcommand{\eff}{\text{eff}}
\newcommand{\Li}{\text{Li}}

\newcommand{\pole}{\text{pole}}
\newcommand{\as}{\alpha_s}
\renewcommand{\to}{\rightarrow}

\newcommand{\sh}{\hat{s}}
\newcommand{\ssh}{\sqrt{\hat{s}}}

\def\Journal#1#2#3#4{{#1} {\bf #2}, #3 (#4)}

\def\NPB{{\em Nucl. Phys.} B}
\def\PLB{{\em Phys. Lett.}  B}
\def\PRL{\em Phys. Rev. Lett.}
\def\PRD{{\em Phys. Rev.} D}

\begin{document}
\thispagestyle{empty}

\preprint{
    \noindent
    \hfill
    \begin{minipage}[t]{6in}
        \begin{flushright}
            BUTP--02/11     \\
            hep-ph/0209006  \\
            \vspace*{1.0cm}
        \end{flushright}
    \end{minipage}
}

\draft

\title{NNLL corrections to the angular distribution and to the 
forward-backward asymmetries in $b \to X_s \ell^+ \ell^-$
\footnote{Work partially supported by Schweizerischer
Nationalfonds, SCOPES and NFSAT (CRDF) programs.}}
\vspace{2.0cm}
\author{H.M. Asatrian$^a$, K. Bieri$^b$, C. Greub$^b$
and A. Hovhannisyan$^a$}
\vspace{2.0cm}
\address{
    a) Yerevan Physics Institute, 2 Alikhanyan Br.,
    375036 Yerevan, Armenia; \\
    b) Institut f\"ur Theoretische Physik, Universit\"at Bern, \\
    CH--3012 Bern, Switzerland.
}
\maketitle
\thispagestyle{empty}
\setcounter{page}{0}

\vspace*{1truecm}
%%
%% ---------- Abstract ----------
%%
\begin{abstract}
We present next-to-next-to leading logarithmic (NNLL) results
for the double differential decay width 
$d\Gamma(b \to X_s \ell^+ \ell^-)/(d\sh \, d\cos(\theta))$, where
$s=\sh \, m_b^2$ is the invariant mass squared of the lepton pair and
$\theta$ is the angle between the momenta of the $b$-quark and the
$\ell^+$, measured in the rest-frame of the lepton pair. From these
results we also derive NNLL results for the lepton forward-backward 
asymmetries, as these quantities are known to be very sensitive to new physics.
While the principal steps in the calculation of the double
differential decay width are the same as
for $d\Gamma(b \to X_s \ell^+ \ell^-)/d\sh$, which is already
known to NNLL precision, genuinely new calculations for
the combined virtual- and gluon bremsstrahlung corrections associated
with the operators $O_7$, $O_9$ and $O_{10}$ are necessary.
In this paper, we neglected certain other bremsstrahlung contributions,
which are known to have only a small impact on 
$d\Gamma(b \to X_s \ell^+ \ell^-)/d\sh$.
We find that the NNLL corrections drastically reduce the renormalization
scale ($\mu$) dependence of the forward-backward asymmetries. 
In particular, $\sh_0$, the position at which the
forward-backward asymmetries vanish, is essentially free of uncertainties
due to the renormalization scale at NNLL precision. We find
$\sh_0^{\rm{NNLL}}=0.162 \pm 0.005$, where the error is dominated
by the uncertainty in $m_c/m_b$. This is to be compared with
$\sh_0^{\rm{NLL}}=0.144 \pm 0.020$, where the error is dominated by
uncertainties due to the choice of $\mu$.
\end{abstract}
\vfill
\setlength{\parskip}{1.2ex}
\newpage
\section{Introduction}
\label{sec:intro}
Rare $B$-meson decays are known to be important sources for informations on
the standard model (SM) of electroweak and strong interactions and its
extensions.  Being very sensitive to the actual physics at the scales
of several hundred GeV, they can be used to distinguish between
different models of fundamental physics and, in particular, to find
significant deviations from the SM predictions. Even restricting
the consideration to the SM case, these decays can be used to retrieve
important information on the properties of the top quark, e.g. to
determine the elements $V_{ts}$ and $V_{td}$ of the 
Cabibbo-Kobayashi-Maskawa (CKM) matrix.

The first measured rare $B$-meson decay was
the exclusive channel $B\to K^*\gamma$, observed by the CLEO
collaboration in 1992 \cite{CLEOrare2}. It was followed by the 
observation of the corresponding inclusive mode $B\to X_s\gamma$ 
\cite{CLEOrare1}. The measured decay
rate \cite{CLEOrare1,ALEPH,BELLE,BABAR} and the photon energy spectrum \cite{CLEO01}
for the latter are in good agreement with 
the predictions of the SM \cite{spektrum,matching,Higgsgiudice,Chetyrkin,matel,matelnew,Gambino01}.
Thus, these observables are well suited for 
constraining the SM extensions, such as two-Higgs doublet models
\cite{BG98,Higgsgiudice,Asatrian3}, left-right symmetric models 
\cite{asatrian3}, supersymmetric models 
\cite{Bertolini91,Giudice99,Misiaksusy,BGHW00,Asatrian:1999sg,BGH01}, etc..

Among the other rare transitions, the inclusive decay $B\to X_s\ell^+\ell^-$
plays a remarkable role. The measurement of various kinematical 
distributions of the decay products
will tighten the constraints on the extensions of the SM or 
perhaps even reveal some deviations, in particular when
combined with improved data on $B\to X_s\gamma$ \cite{ALGH}.

Recently, the BELLE collaboration has reported the observation of the 
exclusive transition $B\to K\mu^+\mu^-$ \cite{Abe:2001dh},
with a rate consistent with the SM predictions. This measurement was 
confirmed by the BABAR collaboration \cite{BABARexcl}.
Very recently, also a measurement of the branching ratio
for the inclusive decay  
$B\to X_s\ell^+\ell^-$ was published by the BELLE collaboration 
\cite{kaneko}.

The interest towards {\it inclusive} rare decays is motivated by the fact
that they can be well 
approximated in suitably chosen kinematical ranges by the underlying
$b$-quark decay. The corrections to this 
simple partonic picture, which can be systematically calculated in the 
framework of Heavy Quark Expansion (HQE), manifest themselves as 
power corrections in $1/m_b$ \cite{Bigi1,Falk:1994dh,Bigi2}. 

The main problem of the theoretical description of $B\to X_s\ell^{+}\ell^{-}$
is due to the long-distance contributions
{}from $\bar{c}c$ resonant states. When the invariant mass
$\sqrt{s}$ of the lepton pair is close to the
mass of a resonance, only model-dependent predictions for such long distance
contributions are available today. It is
therefore unclear whether the theoretical uncertainty can be reduced to less
than $\pm 20\%$ when
integrating over these domains \cite{Ligeti:1996yz}.

However, when restricting $\sqrt s$ to a region below the resonances, the long
distance effects are under control. The left-over effects of the
resonances can again be analyzed within the framework HQE and
manifest themselves as $1/m_c$ power corrections. 
All available studies indicate that for the region 
$0.05<\hat{s}=s/m_b^2<0.25$
these non-perturbative effects are
below 10$\%$
\cite{Falk:1994dh,Ali:1997bm,chen:1997,Buchalla:1998ky,Buchalla:1998mt,Krueger:1996}.
Consequently, the differential decay rate for $B \to X_s \ell^+ \ell^-$
can be precisely predicted in this region, using
renormalization group improved perturbation theory.
It was pointed out in the literature that the 
invariant mass distribution of the lepton pair and the
forward-backward asymmetries are particularly sensitive to
new physics in this kinematical window \cite{Ali:1997bm,Ball,Lunghi,Silvestrini,Asatrian3}.

Although the consideration of inclusive decays allows to avoid the most 
difficult issues of hadronic physics, the {\it perturbative} 
QCD corrections play a very important role for
all rare $B$-decays. Calculations of the
next-to-leading logarithmic (NLL)  QCD corrections to the invariant mass 
distribution of the lepton pair ($d\Gamma(b \to X_s \ell^+ \ell^-)/d\sh$)
were performed in refs. \cite{Misiak:1993bc} and \cite{Buras:1995dj}.
It turned out that the NLL result suffers from a relatively large
($\pm 16\%$) dependence on the matching scale $\mu_W$.
To reduce it, next-to-next-to leading logarithmic (NNLL)
corrections to the Wilson coefficients were 
calculated by Bobeth et al. \cite{Bobeth:2000mk}.
This required a two-loop matching calculation of the full SM theory
onto the effective theory,
followed by a renormalization group evolution of the
Wilson coefficients, using up to three-loop anomalous dimensions
\cite{Bobeth:2000mk,Chetyrkin}.
Including these NNLL corrections to the Wilson coefficients, the
matching scale dependence is indeed removed to a large extent.

As pointed out in ref. \cite{Bobeth:2000mk}, this partial NNLL result
suffers from a relatively large ($\sim \pm 13\%$)
renormalization scale ($\mu_b$) dependence ($\mu_b \sim \Order(m_b)$).
In order to further improve the theoretical prediction, we recently calculated
the virtual two-loop corrections to the matrix elements  $\bra  s\, \ell^+ 
\ell^-|O_i|b \ket$ 
($i=1,2$) as well as the virtual ~$(\alpha_s)$ one-loop corrections to
$O_7$,...,~$O_{10}$ and the corresponding bremsstrahlung corrections 
\cite{Asatrian:2001de,Asatrian1,Asatrian2}.  This improvement reduced 
the renormalization scale dependence of $d\Gamma(b \to X_s \ell^+ \ell^-)/d\sh$
by a factor of 2.

In the present paper, we present a calculation of the double 
differential decay width $d\Gamma$/($d\sh \, d\cos(\theta))$ and the 
forward-backward asymmetries for the 
decay $b  \to X_s \ell^+ \ell^-$ at NNLL precision.
$\theta$ denotes the angle between the momenta of the positively 
charged lepton ($\ell^+$) and the $b$-quark, measured in the rest-frame of 
the lepton pair. It is well-known that the measurement of the forward-backward
asymmetries along with detailed experimental information on the invariant
mass distribution of the lepton pair can be used, in combination
with the measurement of the radiative decay $B \to X_s \gamma$, 
to perform ``a model-independent test'' 
of the SM \cite{goto,ali1,ALGH}. 
In particular,  for some extensions of the SM the branching ratio for
the process $B \to X_s \gamma$ is the same as in the SM,  
but the Wilson coefficient $C_7$ has opposite sign
\cite{ALGH,asatrian4,asatrian5,Asatrian:1999sg}. 
As shown in refs. \cite{goto,ali1,ALGH}, 
the measurement of the shape of the forward-backward asymmetries as a function
of $\sh$ in the process $B \to X_s \ell^+ \ell^-$
would allow to determine whether the SM sign or the opposite sign
of $C_7$ is realized in nature. Needless to say, the measurement of the 
forward-backward asymmetries also yields 
additional (and complementary)
information for determining the Wilson coefficients  
$C_9$ and $C_{10}$. 

Being a crucial observable in the search for new physics in rare
$B$ decays, the forward-backward asymmetries should be calculated
in the SM as precisely as possible. As the available NLL results
suffer from a large dependence on the renormalization scale, we
perform a NNLL calculation of these asymmetries in the present paper.
Note that the NNLL corrections to the 
forward-backward asymmetries cannot be straightforwardly derived from our
previous results for $d\Gamma(b \to X_s \ell^+ \ell^-)/d\sh$, i.e., 
a partial recalculation is required.
In particular, this concerns the bremsstrahlung contributions 
associated with the operators $O_7$, $O_9$ and $O_{10}$, which are
needed for the cancellation of the infrared- and collinear singularities 
in the virtual corrections.

The paper is organized as follows: In section \ref{sec:two} we recall
the theoretical framework. Section \ref{sec:three} is devoted to the
previous results on $d\Gamma(b \to X_s \ell^+ \ell^-)/d\sh$ and explains why
modifications are needed for the derivation of the double differential
decay width. In section \ref{sec:four} the analytical
results for the double differential
decay width and for the forward-backward asymmetries are presented. In
sections \ref{sec:five}, \ref{sec:six} and \ref{sec:seven} the technical
issues needed for the derivation of the double differential decay width
are explained. In section \ref{sec:eight} a detailed
phenomenological analysis for the forward-backward asymmetries is presented;
the angular distributions are also shortly discussed. Finally,
in section \ref{sec:nine} we briefly summarize our paper. 
In this section
we also compare our results on the forward-backward asymmetries with
those reported in ref. \cite{Isidori}, which appeared
when we were working out the double differential decay width.

\section{Theoretical framework}
\label{sec:two}
As mentioned above, the QCD corrections give significant 
(sometimes even dominant) contributions to the decay rates of rare
processes. 
The most efficient tool for analyzing these corrections
in a systematic way is the effective Hamiltonian 
technique. The effective Hamiltonian for a particular decay channel of
a $b$-quark is obtained by integrating out the heavy degrees
of freedom which are (in the context of the SM) the 
top quark, the $W^{\pm}$ and $Z^0$
bosons. The effective Hamiltonian for the decay $b\to X_s \ell^+\ell^-$ 
reads
\begin{eqnarray}
    \label{Heff}
    {\cal H}_{\text{eff}} =  - \frac{4G_F}{\sqrt{2}} V_{ts}^* V_{tb}
    \sum_{i=1}^{10} C_i \, O_i,
\end{eqnarray}
where we have omitted the contributions which are weighed 
by the small CKM factor $V_{us}^* V_{ub}$.
The dimension six effective operators can be chosen as 
\cite{Bobeth:2000mk}
\begin{equation}
\label{oper}
\begin{array}{rclrcl}
    O_1    & = & (\bar{s}_{L}\gamma_{\mu} T^a c_{L })
                (\bar{c}_{L }\gamma^{\mu} T^a b_{L})\, , &
    O_2    & = & (\bar{s}_{L}\gamma_{\mu}  c_{L })
                (\bar{c}_{L }\gamma^{\mu} b_{L})\, ,\\ \vspace{0.2cm}
    O_3    & = & (\bar{s}_{L}\gamma_{\mu}  b_{L })
                \sum\limits_q (\bar{q}\gamma^{\mu}  q)\, , &
    O_4    & = & (\bar{s}_{L}\gamma_{\mu} T^a b_{L })
                \sum\limits_q (\bar{q}\gamma^{\mu} T^a q)\, , \\ \vspace{0.2cm}
    O_5    & = & \bar{s}_L \gamma_{\mu} \gamma_{\nu}
                \gamma_{\rho}b_L
                \sum\limits_{q}\bar{q} 
\gamma^{\mu} \gamma^{\nu}\gamma^{\rho}q\, , &
    O_6    & = & \bar{s}_L \gamma_{\mu} \gamma_{\nu}
                \gamma_{\rho} T^a b_L
                \sum\limits_{q}\bar{q} \gamma^{\mu} \gamma^{\nu}
                \gamma^{\rho} T^a q\, ,    \vspace{0.2cm} \\ \vspace{0.2cm}
    O_7    & = & \frac{e}{g_s^2} m_b (\bar{s}_{L} \sigma^{\mu\nu}
                b_{R}) F_{\mu\nu}\, , &
    O_8    & = & \frac{1}{g_s} m_b (\bar{s}_{L} \sigma^{\mu\nu}
                T^a b_{R}) G_{\mu\nu}^a\, , \\ \vspace{0.2cm}
    O_9    & = & \frac{e^2}{g_s^2}(\bar{s}_L\gamma_{\mu} b_L)
               (\bar{\ell}\gamma^{\mu}\ell)\, , &
    O_{10} & = & \frac{e^2}{g_s^2}(\bar{s}_L\gamma_{\mu} b_L)
                (\bar{\ell}\gamma^{\mu} \gamma_{5} \ell)\, .
\end{array}
\end{equation}
The subscripts $L$ and $R$ refer to left- and right-handed fermion fields. 
The factors $1/g_s^2$ in the definition
of the operators $O_7$, $O_9$ and $O_{10}$, as well as the factor
$1/g_s$ present in $O_8$ have been chosen by Misiak \cite{Misiak:1993bc}
in order to simplify
the organization of the calculation: With these definitions,
the one-loop anomalous dimensions (needed for a leading logarithmic
(LL) calculation) of the operators $O_i$ are all proportional to $g_s^2$,
while
two-loop anomalous dimensions (needed for a next-to-leading logarithmic
(NLL) calculation) are proportional to $g_s^4$, etc..

In this setup, the principal steps
which lead to a (formally) LL, NLL, NNLL prediction for the decay amplitude
for $b \to X_s \ell^+ \ell^-$ are the following:
\begin{enumerate}
\item
A matching calculation between the full SM theory and the effective
theory has to be performed
in order to determine the Wilson coefficients $C_i$
at the high scale $\mu_W\sim m_W,m_t$. At this scale, the coefficients
can be worked out in fixed order perturbation theory, i.e. they can be expanded
in $g_s^2$:
\begin{equation}
    C_i(\mu_W) = C_i^{(0)}(\mu_W)
    + \frac{g_s^2}{16\pi^2} C_i^{(1)}(\mu_W)
    + \frac{g_s^4}{(16\pi^2)^2} C_i^{(2)}(\mu_W) + O(g_s^6) \, .
\end{equation}
At LL order, only $C_i^{(0)}$ is needed, at NLL order also $C_i^{(1)}$,
etc.. While the coefficient $C_7^{(2)}$, which is needed for
a NNLL analysis, is known for quite some time \cite{matching}, 
$C_{9}^{(2)}$ and  $C_{10}^{(2)}$ have been
calculated only recently \cite{Bobeth:2000mk} 
(see also \cite{Buchalla:1999ba}).
\item
The renormalization group equation (RGE) has to be solved in order
to get the Wilson
coefficients at the low scale $\mu_b \sim m_b$.
For this RGE step
the anomalous dimension matrix to the relevant
order in $g_s$ is required, as described above.
After these two steps one can decompose the Wilson coefficients
$C_i(\mu_b)$ into a LL, NLL and NNLL part according to
\begin{equation}
    \label{wilsondecomplow}
    C_i(\mu_b) = C_i^{(0)}(\mu_b)
    + \frac{g_s^2(\mu_b)}{16\pi^2} C_i^{(1)}(\mu_b)
    + \frac{g_s^4(\mu_b)}{(16\pi^2)^2} C_i^{(2)}(\mu_b) + O(g_s^6) \, .
\end{equation}
\item
In order to get the decay amplitude,
the matrix elements $\bra s \ell^+ \ell^-|O_i(\mu_b)|b \ket$
have to be calculated. At LL precision, only the operator $O_9$
contributes, as this operator is the only one which at the same time has
a Wilson coefficient starting at lowest order and an explicit
$1/g_s^2$ factor in the definition. Hence, in the NLL precision
QCD corrections (virtual and bremsstrahlung)
to the matrix element of $O_9$ are needed. They have been calculated
a few years ago \cite{Misiak:1993bc,Buras:1995dj}. At NLL precision, also
the other operators start contributing,
viz. $O_7(\mu_b)$ and
$O_{10}(\mu_b)$ contribute at tree-level
and the four-quark operators $O_1,...,O_6$
at one-loop level. Accordingly,
QCD corrections to the latter matrix elements
are needed for a NNLL prediction of the decay amplitude.
\end{enumerate}
As known for a long time \cite{Grinstein:1989}, the formally leading term
$\sim (1/g_s^2) C_9^{(0)}(\mu_b)$ to the amplitude
for $b \to s \ell^+ \ell^-$  is smaller than
the NLL term $\sim (1/g_s^2) [g_s^2/(16 \pi^2)] \, C_9^{(1)}(\mu_b)$.
As in our earlier papers on the NNLL prediction for 
$\mbox{BR}(b \to X_s \ell^+ \ell^-)$ 
\cite{Asatrian:2001de,Asatrian1,Asatrian2},
we adapt our systematics to the numerical situation
and treat the sum of these two terms as a NLL contribution.
This is, admittedly, some abuse of language, because the decay
amplitude then starts with a term which is called NLL. Using this
adapted counting, no QCD corrections to the matrix elements
$\bra s \ell^+ \ell^-|O_i(\mu_b)|b \ket$ ($i=1,...,10$)
are needed when working at NLL precision,
while one-gluon (virtual- and bremsstrahlung) corrections are 
necessary at NNLL precision. 
%
%It is necessary to evaluate also the  corresponding bremsstrahlung 
%corrections which are necessary, in particular, for the 
%cancellation of the infrared and 
%collinear singularities from the virtual corrections.  
%This was done for the case
%of the differential decay width $d\Gamma$/$ds$ in our previous papers. 
%However for  the double differential decay width  
%$d\Gamma$/$dsd\cos(\theta)$ this problem has not been solved yet 
%at the NNLL level.

\noindent
When working out 
in the following the QCD corrections to the matrix elements, 
we often also use the related operators 
$\widetilde{O}_7$,..., $\widetilde{O}_{10}$, defined according to
\begin{equation}
    \widetilde{O}_j = \frac{\alpha_s}{4\,\pi} \, O_j \, \, ,
    \quad (j=7,...,10) \, ,
\end{equation}
with the corresponding Wilson coefficients
\begin{equation}
    \widetilde{C}_j = \frac{4\,\pi}{\alpha_s} \, C_j \, \, , \quad 
(j=7,...,10) \, .
\end{equation}
\section{Previous results for 
$\boldsymbol{\symbol{100}\Gamma/\symbol{100}\sh}$
and modifications needed for 
$\boldsymbol{\symbol{100}\Gamma/(\symbol{100}\sh \, \symbol{100}\cos \theta)}$}
\label{sec:three}
To obtain the NNLL approximation for 
$d\Gamma(b \to X_s \ell^+ \ell^-)/d\sh$,
using the modified counting discussed above,
virtual- and gluon bremsstrahlung corrections 
were calculated in refs. \cite{Asatrian:2001de,Asatrian1,Asatrian2}
and combined with the Wilson coefficients evaluated to
the corresponding precision.
For completeness, we briefly repeat these results, and
put them into a slightly different form than presented in
refs. \cite{Asatrian:2001de,Asatrian1,Asatrian2}. The 
distribution of the invariant mass squared of the lepton pair
can be written as
\bea
\label{rarewidth}
    && \frac{d\Gamma(b\to X_s\, \ell^+\ell^-)}{d\sh} =
    \left(\frac{\alpha_{\text{em}}}{4\,\pi}\right)^2
    \frac{G_F^2\, m_{b,\pole}^5\left|V_{ts}^*V_{tb}^{}\right|^2} 
     {48\,\pi^3}(1-\sh)^2 \times \nonumber \\ 
    && \hspace{1cm}
    \left\{ \left( 1+2\,\sh \right) \left 
     (\left |\widetilde C_9^{\eff}\right |^2 +
    \left |\widetilde C_{10}^{\eff}\right |^2 \right) \, 
    [1+\frac{2 \, \alpha_s}{\pi} \, \omega_{99}(\sh)] 
 + 4\left(1+\frac{2}{\sh}\right)\left |\widetilde C_7^{\eff}\right |^2
    [1+\frac{2 \, \alpha_s}{\pi} \, \omega_{77}(\sh)] 
\right. \nonumber \\
 && \hspace{1cm} \left.
 + 12\, \mbox{Re}\left (\widetilde C_7^{\eff} \widetilde C_9^{\eff*}\right )
    [1+\frac{2 \, \alpha_s}{\pi} \, \omega_{79}(\sh)]   \right\} 
+ \frac{d\Gamma^{\rm{Brems,A}}}{d\sh} + \frac{d\Gamma^{\rm{Brems,B}}}{d\sh}
\, .
\eea
$\frac{d\Gamma^{\rm{Brems,A}}}{d\sh}$ and $\frac{d\Gamma^{\rm{Brems,B}}}{d\sh}$
are the finite bremsstrahlung corrections discussed in detail
in ref. \cite{Asatrian2} (see eqs. (13) and (22) in this reference).
The other bremsstrahlung corrections, associated with the operators
$\wtO_7$, $\wtO_9$ and $\wtO_{10}$ suffer from infrared- and
collinear singularities. They are contained, combined with the corresponding
virtual corrections, in the quantities 
$\omega_{99}(\sh)$,  $\omega_{77}(\sh)$ and $\omega_{79}(\sh)$.
As they will be needed in the construction of the double
differential decay width, we repeat their explicit form in appendix A. 
The virtual corrections to the matrix elements of $O_1$, $O_2$
and $O_8$, on the other hand, are infrared finite. They can be written 
as multiples of tree-level matrix elements of the operators
$\wtO_7$, $\wtO_9$ and $\wtO_{10}$, and are usually absorbed 
(through the functions $F_i^{(j)}$ ($i=1,2,8; j=7,9$)) 
into the effective Wilson coefficients
$\widetilde{C}_7^{\eff}$, 
$\widetilde{C}_9^{\eff}$ and
$\widetilde{C}_{10}^{\eff}$, which read 
\bea
    \label{effcoeff7}
    \widetilde C_7^{\eff} &=& A_7 
        -\frac{\alpha_{s}(\mu)}{4\,\pi}\left(C_1^{(0)} F_1^{(7)}(\s)+
        C_2^{(0)} F_2^{(7)}(\s) + A_8^{(0)} F_8^{(7)}(\s)\right),  \\
    \label{effcoeff9}
    \widetilde C_9^{\eff} &=& A_9 + T_9 \, h (\hat m_c^2, \s)+U_9 \, h (1,\s) +
        W_9 \, h (0,\s) \nn \\
        && -\frac{\alpha_{s}(\mu)}{4\,\pi}\left(C_1^{(0)} F_1^{(9)}(\s) + 
       C_2^{(0)} F_2^{(9)}(\s)+
        A_8^{(0)} F_8^{(9)}(\s)\right),  \\
    \label{effcoeff10}
    \widetilde C_{10}^{\eff} &=& A_{10}.
\eea
The quantities $C_1^{(0)}$, $C_2^{(0)}$, $A_7$, $A_8^{(0)}$, $A_9$, $A_{10}$, 
$T_9$, $U_9$ and $W_9$ are Wilson
coefficients or linear combinations thereof. Their analytical 
expressions and numerical values are given in appendix B.
The one-loop function $h (\hat m_c^2,\hat{s})$ is also given there,
while the two-loop functions
$F_{1,2}^{(7),(9)}$, and the one-loop functions $F_8^{(7),(9)}$
are given in ref.~\cite{Asatrian1}. 
We remind the reader that in the above results the QCD corrections
to the matrix elements of the operators $O_3-O_6$ were not taken
into account {\it systematically}, as they are weighted by small
Wilson coefficients.

It may appear as a surprise that a NNLL calculation 
for $d\Gamma(b \to X_s \ell^+ \ell^-)/d\sh$ is available,
while the corresponding result for 
$d^2\Gamma(b \to X_s \ell^+ \ell^-)/(d\sh \, d\cos(\theta))$ is
still missing. The reason is a technical one. When aiming only at
$d\Gamma(b \to X_s \ell^+ \ell^-)/d\sh$, it is convenient to 
integrate in a first
step over the lepton momenta after multiplying the well-known
expression for the fully differential decay width by a factor $1$ in the
form (note that $\sh=q^2/m_b^2$)
\begin{equation}
\label{deltainsertion}
1 = \int \delta^d (q - l_1 - l_2) \, d^dq \, .
\end{equation}
This is precisely what we did in our previous works
\cite{Asatrian:2001de,Asatrian1,Asatrian2}. It is evident that
after this step the angular correlation between hadronic and leptonic
variables is lost. For this reason, the phase space integrations
have to be done in another way when aiming at a calculation
of the double differential decay width. While these modifications 
connected to phase space are straightforward for the lowest order
and the virtual corrections, where only three particles are in the final
state, a genuinely new calculation is needed for the gluon bremsstrahlung
process with four particles in the final state.

We decide to postpone the discussion of these technical issues
to sections \ref{sec:five}--\ref{sec:seven}, as we prefer to first present 
the final results for the double differential decay width and
for the forward-backward asymmetries. 
%
%For the moment, 
%we only want to stress that 
%$\theta$ is the angle between $\vec{l^+}$ and 
%$\vec{p}_b$ measured in the {\it rest frame of the lepton pair}.
%Therefore, it turns out to be convenient
%to first derive an expression for the
%fully differential decay width in this frame. 
%
\section{NNLL results for the double differential decay width
and the forward-backward asymmetries}
\label{sec:four}
We write the double differential decay width 
$d^2\Gamma(b \to X_s \ell^+ \ell^-)/(d\sh \, dz)$ ($z=\cos(\theta)$)
in a form which is analogous to the expression for
$d\Gamma(b \to X_s \ell^+ \ell^-)/d\sh$ in eq. (\ref{rarewidth}).
We obtain
\bea
\label{doublewidth}
    \frac{d^2\Gamma(b\to X_s\, \ell^+\ell^-)}{d\sh \, dz} = &&
    \left(\frac{\alpha_{\text{em}}}{4\,\pi}\right)^2
    \frac{G_F^2\, m_{b,\pole}^5\left|V_{ts}^*V_{tb}^{}\right|^2}
    {48\,\pi^3}(1-\sh)^2 \nn \\
    &&
    \times \left\{ \frac{3}{4} [(1-z^2)+\sh (1+z^2)] \, 
    \left( \left |\widetilde C_9^{\eff}\right |^2 +
    \left |\widetilde C_{10}^{\eff}\right |^2 \right) \, 
    \left( 1+\frac{2 \as}{\pi} \, f_{99}(\sh,z) \right) \right. \nn  \\
    && +
    \frac{3}{\sh} [(1+z^2)+\sh(1-z^2)] 
    \, \left | \widetilde C_7^{\eff}\right |^2 
    \left( 1+\frac{2 \as}{\pi} \, f_{77}(\sh,z) \right)  \nn  \\
    && - 3 \, \sh \, z \, \mbox{Re}(\widetilde C_9^{\eff} \widetilde 
    C_{10}^{\eff *})
    \,    \left( 1+\frac{2 \as}{\pi} \, f_{910}(\sh) \right)  \nn  \\
    && + 6  \, \mbox{Re}(\widetilde C_7^{\eff} \widetilde C_9^{\eff *})
    \,    \left( 1+\frac{2 \as}{\pi} \, f_{79}(\sh,z) \right)  \nn  \\ 
   && \left. - 6 \, z \, \mbox{Re}(\widetilde C_7^{\eff} 
    \widetilde C_{10}^{\eff *})
    \,    \left( 1+\frac{2 \as}{\pi} \, f_{710}(\sh) \right) \right\} \, .
\eea
The effective Wilson coefficients are the same as those used for
$d\Gamma/d\sh$; they are given
in eqs. 
(\ref{effcoeff7})-(\ref{effcoeff10}). In particular, they contain
the virtual corrections to the matrix elements of the operators
$O_1$, $O_2$ and $O_8$. The sum of virtual- and bremsstrahlung corrections
to the matrix elements of $O_7$, $O_9$ and $O_{10}$ is incorporated
in the functions $f_{99}(\sh,z)$, $f_{77}(\sh,z)$, $f_{910}(\sh)$,  
$f_{79}(\sh,z)$ and $f_{710}(\sh)$. These functions are the analogues of 
$\omega_{99}(\sh)$, $\omega_{77}(\sh)$ and $\omega_{79}(\sh)$ which
enter eq. (\ref{rarewidth}). 
As indicated in the notation, the functions $f_{710}$ and $f_{910}$
only depend on $\sh$, while $f_{99}$, $f_{77}$ and $f_{79}$ depend
also on $z$. In eq. (\ref{doublewidth}) we do not include the 
purely finite bremsstrahlung corrections, which in the case for
$d\Gamma/d\sh$ were encoded in eq. (\ref{rarewidth}) in the last two
terms. This omission is motivated by the fact that these corrections have a
negligible impact on $d\Gamma/d\sh$.

We now turn to the forward-backward asymmetries. We will investigate
both, the so-called normalized- and the unnormalized forward-backward
asymmetry. The normalized version, $\overline{A}_{\text{FB}}(\sh)$, is defined
as
\bea
\label{asymmnorm}
   \overline{A}_{\text{FB}}(\sh) &=& \frac{
    \int_{-1}^1 \frac{d^2\Gamma(b\to X_s\, \ell^+\ell^-)}{d\sh \, dz} 
    \, \mbox{sgn}(z) \, dz}{
    \int_{-1}^1 \frac{d^2\Gamma(b\to X_s\, \ell^+\ell^-)}{d\sh \, dz} 
     \, dz} \, ,
\eea
while the definition of the unnormalized forward-backward asymmetry 
$A_{\text{FB}}(\sh)$ reads
\bea
\label{asymmunnorm}
   A_{\text{FB}}(\sh) &=& \frac{
    \int_{-1}^1 \frac{d^2\Gamma(b\to X_s\, \ell^+\ell^-)}{d\sh \, dz} 
    \, \mbox{sgn}(z) \, dz}{\Gamma(B \to X_c e \bar{\nu}_e)} \, 
    \mbox{BR}_{\rm{sl}}  \, .
\eea
The denominator in eq. (\ref{asymmunnorm}) is the semileptonic
decay width, which is usually put into the definition of the
unnormalized forward-backward asymmetry in order to cancel
the fifth power of $m_b$ present in the numerator. The expression
for $\Gamma(b \to X_c e \bar{\nu}_e)$ is well-known, including
$O(\alpha_s)$ QCD corrections \cite{Nir}, and can be taken e.g.
form ref. \cite{Asatrian1}. The factor $\mbox{BR}_{\rm{sl}}$ 
in eq. (\ref{asymmunnorm}) denotes the measured semileptonic
branching ratio of the $B$-meson.

\begin{figure}[t]
    \begin{center}
    \leavevmode
    \includegraphics[height=8cm]{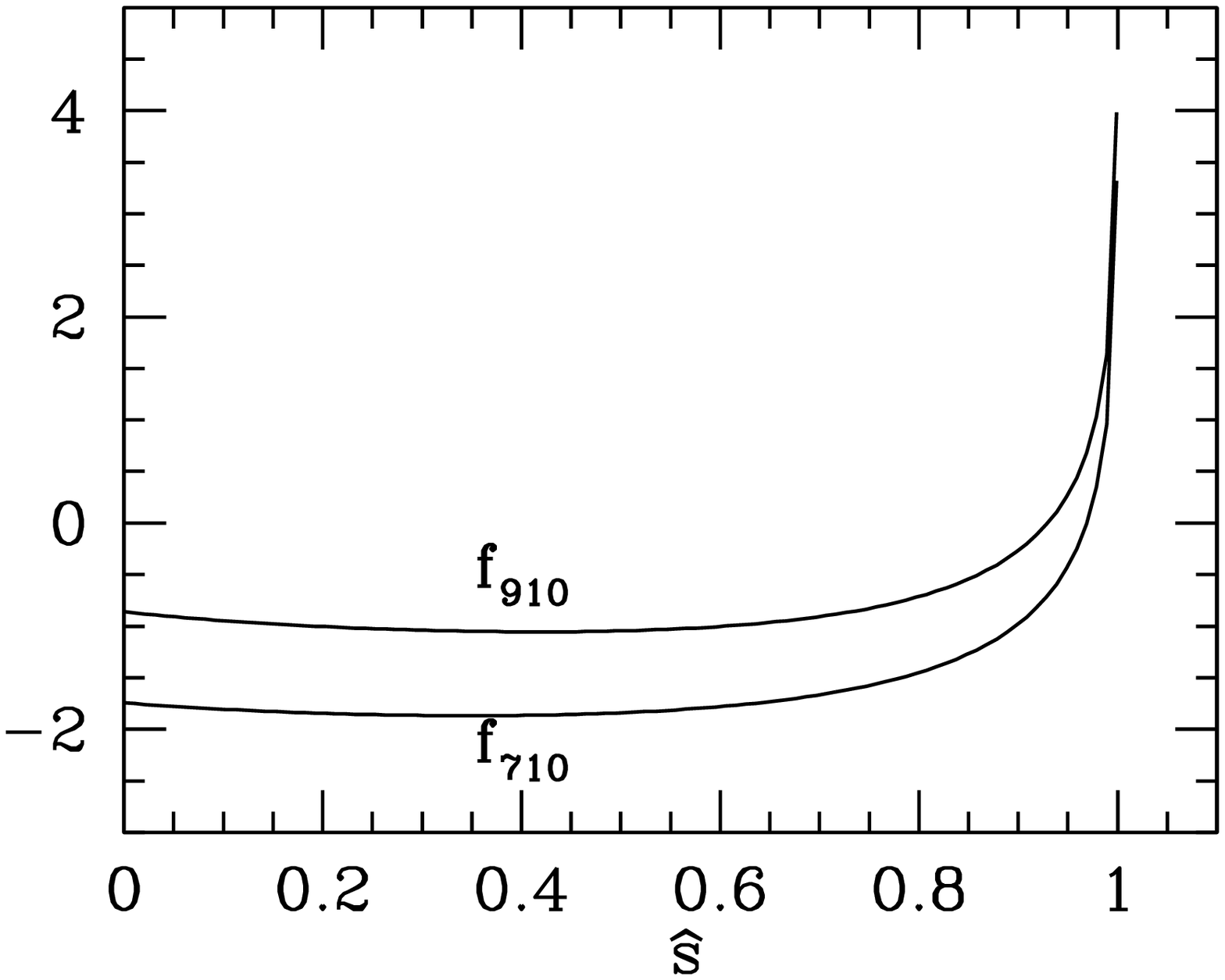}
    \vspace{2ex}
    \caption{Functions $f_{710}(\sh)$ and $f_{910}(\sh)$ 
    which in the forward-backward asymmetries incorporate virtual- 
    and bremsstrahlung corrections
    to the ($O_7,O_{10}$) and ($O_9,O_{10}$) interference terms. 
    $\mu/m_b$=1.}
    \label{fig:9}
    \end{center}
\end{figure}

\noindent
In the numerator, both asymmetries involve the same forward-backward
integral over the double differential decay width. For this integral
one obtains
\bea
\label{asymmint}
   && \int_{-1}^1 \frac{d^2\Gamma(b\to X_s\, \ell^+\ell^-)}{d\sh \, dz} 
    \, \mbox{sgn}(z) \, dz = 
    \left(\frac{\alpha_{\text{em}}}{4\,\pi}\right)^2
    \frac{G_F^2\, m_{b,\pole}^5\left|V_{ts}^*V_{tb}^{}\right|^2}
    {48\,\pi^3}(1-\sh)^2 \nn \\
    &&
    \times \left[
    - 3 \, \sh \, \mbox{Re}(\widetilde C_9^{\eff} \widetilde 
    C_{10}^{\eff *})
    \,    \left( 1+\frac{2 \as}{\pi} \, f_{910}(\sh) \right)  
    - 6  \, \mbox{Re}(\widetilde C_7^{\eff} 
    \widetilde C_{10}^{\eff *})
    \,    \left( 1+\frac{2 \as}{\pi} \, f_{710}(\sh) \right) \right] \, .
\eea
This result shows that only the interference terms
($O_9,O_{10}$) and ($O_7,O_{10}$) contribute to
the asymmetries. 
The two functions $f_{710}(\sh)$ and $f_{910}(\sh)$ in eq. (\ref{asymmint}),
which incorporate the sum of virtual- and bremsstrahlung corrections
to the matrix elements of $O_7$, $O_9$ and $O_{10}$,
are plotted in fig. \ref{fig:9}.
%---------- figure ----------

The main new result
of this paper is encoded in the functions
$f_{99}(\sh,z)$, $f_{77}(\sh,z)$, $f_{910}(\sh)$, 
$f_{79}(\sh,z)$, and $f_{710}(\sh)$, which we managed
to calculate analytically. We obtain ($\mu$ denotes the renormalization scale)
\bea
  \label{eq:f710}
  f_{710} & = & -\frac{1}{18\sh
    (1-\sh)^2}\left(6\sh(3+9\sh-2\sh^2)\text{Li}_2(\sh)-
   12\sh(1+13\sh-4\sh^2)\text{Li}_2(\ssh)\right.
  \nn \\
& &
  +3(1-23\sh+23\sh^2-\sh^3)\ln(1-\sh)+6\sh(13-16\sh+3\sh^2)\ln(1-\ssh)
  \nn \\
& &
  \left.+\sh\big(5\pi^2(1+\sh)-3(5-20\ssh+\sh)(1-\ssh)^2\big)+24\sh(1-\sh)^2
\ln\left(\mu/m_b\right)\right) \, ,
\eea
\bea
  \label{eq:f910}
f_{910} & = &
  -\frac{1}{9\sh(1-\sh)^2}\left(6\sh(1+3\sh-\sh^2)\text{Li}_2(\sh)-12\sh^2
(5-2\sh)\text{Li}_2(\ssh)\right.
  \nn \\
& & +3(1-10\sh+11\sh^2-2\sh^3)\ln(1-\sh)+6\sh(5-7\sh+2\sh^2)\ln(1-\ssh) \nn \\
& & +\left.\sh\big(3(4\ssh-3)(1-\ssh)^2+\pi^2(2+\sh)\big)\right) \, , 
\eea
\bea
  \label{eq:f79}
f_{79} & = &
  -\frac{1}{36\sh(1-\sh)^2}\bigg(3\sh(1+\ssh)^2\left(3(5+z^2)-3\ssh(11-z^2)+
16\sh\right)\text{Li}_2(\sh)
  \nn \\
& &
  +12\sh\ssh(3+\sh)(1-3z^2)\text{Li}_2(\ssh)+3(1-\sh)^2\left(3-z^2+
\sh(9+z^2)\right)\ln(1-\sh)
  \nn \\
& &
  +3\sh^2\left(13-15z^2-\sh(5+z^2)\right)\ln(\sh)+3\sh\left(7+3z^2+8\sh-
\ssh(17-3z^2)\right)
  \nn \\
& & \times
  (1+\ssh)^2\ln(1-\sh)\ln(\sh)+6\sh\ssh(3+\sh)(1-3z^2)\ln(1-\ssh)\ln(\sh) \nn
  \\
& & -6\sh(1-\sh)\left(5z^2-\sh(4-3z^2)\right)+\sh\pi^2\left(7+3z^2+8\sh^2-
\sh(19-9z^2)\right)
  \nn \\
&  & +48\sh(1-\sh)^2\ln\left(\mu/m_b\right)\bigg) \, ,
\eea
\bea
  \label{eq:f77}
f_{77} & = &
  -\frac{1}{18(1-\sh)^2\left(1+z^2+\sh(1-z^2)\right)}\bigg(12\ssh(3+6\sh-\sh^2)
(1-3z^2)\text{Li}_2(\ssh)
  \nn \\
& &
  +3(1+\ssh)^2\left(8(1+z^2)-\ssh\left(19-14\ssh+15\sh-8\sh\ssh\right.\right.
  \nn \\
& &
  \left.\left.+\left(7-\ssh(2-\ssh)(3+8\ssh)\right)z^2\right)\right)
\text{Li}_2(\sh)+6(1-\sh)^2\left(5+z^2+\sh(1-z^2)\right)
  \nn \\
& & \times
  \ln(1-\sh)+6\sh\left(5-7z^2+\sh(1-11z^2)-2\sh^2(1-z^2)\right)\ln(\sh)+3(1+\ssh)^2
  \nn \\
& & \times
  \left(4(1+z^2)-\ssh\left(11-z^2-\ssh\left(6-7\ssh+4\sh+(2-\ssh)
(3+4\ssh)z^2\right)\right)\right)
  \nn \\
& & \times \ln(1-\sh)\ln(\sh)+6\ssh(3+6\sh-\sh^2)(1-3z^2)\ln(1-\ssh)\ln(\sh)
  \nn \\
& &
  +2\left(2\pi^2\left(1+z^2-3\sh(1-z^2)-\sh^2(1-3z^2)+\sh^3(1-z^2)\right)+
(1-\sh)\left(\sh(19-68z^2)
  \right.\right. \nn \\
& & \left.\left.+4(1+z^2)-\sh^2(11-16z^2)\right)\right)+48(1-\sh)^2
\left(1+z^2+\sh(1-z^2)\right)\ln\left(\mu/m_b\right)\bigg) \, ,
\eea
\bea
  \label{eq:f99}
  f_{99} & = &
  \frac{1}{18(1-\sh)^2\left(1+\sh-z^2(1-\sh)\right)}\bigg(12\ssh(5+12\sh-\sh^2)
(1-3z^2)\text{Li}_2(\ssh)-3(1+\ssh)^2
  \nn \\
& & \times
  \left(8-11\ssh+20\sh-17\sh\ssh+8\sh^2-(1+\ssh)\left(8-\ssh
\left(9-\ssh(21-8\ssh)\right)\right)z^2\right)
  \nn \\
& & \times
  \text{Li}_2(\sh)+6\sh\left(3-13z^2+\sh(9-23z^2)+2\sh^2(1+z^2)\right)
\ln(\sh)-12(1-\sh)^2
  \nn \\
& & \times\left(2-z^2+\sh(1+z^2)\right)\ln(1-\sh)-3(1+\ssh)^2
\left(4-4z^2-\ssh(3+7z^2)+12\sh(1-z^2)\right.
  \nn \\
& &
  \left.-\sh\ssh(9+5z^2)+4\sh^2(1+z^2)\right)\ln(1-\sh)\ln(\sh)+
6\ssh(5+12\sh-\sh^2)(1-3z^2)\ln(\sh)
  \nn \\
& &
  \times \ln(1-\ssh)+3(1-\sh)\left(5-5z^2+\sh(28-66z^2)-\sh^2(5-3z^2)\right)
  \nn \\
& & -2\pi^2\left(2-2z^2+5\sh(1-3z^2)-\sh^2(1+9z^2)+2\sh^3(1+z^2)\right)\bigg)
\, .
\eea

\noindent
In the following three sections, we discuss the technical issues
needed to derive the functions
$f_{99}(\sh,z)$, $f_{77}(\sh,z)$, $f_{910}(\sh)$, 
$f_{79}(\sh,z)$ and $f_{710}(\sh)$.
In section \ref{sec:five} we discuss the regularization of infrared- and
collinear singularities at the level of the matrix elements
(or matrix elements squared). In section \ref{sec:six} we first derive 
a formula for the fully differential decay width in the rest frame
of the lepton pair, which for us was crucial in order to
derive {\it analytical results} for the functions $f$. Using this formula, 
we derive the phase space expressions for the double differential decay width.
Finally, in section \ref{sec:seven} we present some tricks, which allow
us to drastically simplify the calculation of the gluon bremsstrahlung
process. These tricks are based on the {\it universal 
structure of infrared- and collinear singularities}.
\section{Regularization of infrared- and collinear singularities}
\label{sec:five}
\begin{figure}[t]
    \begin{center}
    \leavevmode
    \includegraphics[height=3cm,bb=106 629 529 725]{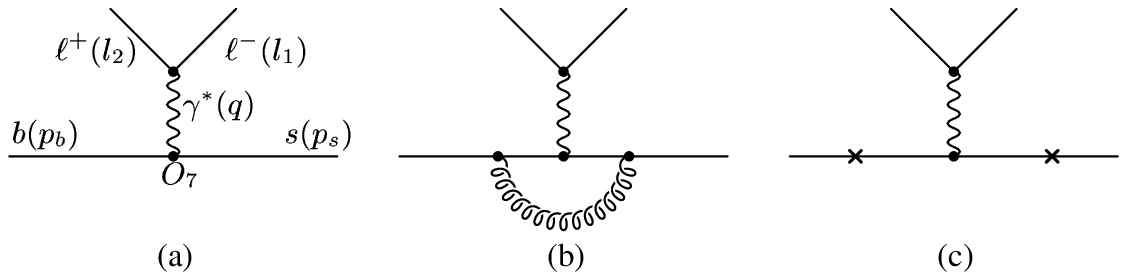}
    \vspace{2ex}
    \caption{Feynman diagrams associated with the operator $O_7$.
    (a) shows the lowest order diagram, (b) and (c) show virtual- and
    bremsstrahlung corrections, respectively. The cross denotes the possible
    emission of the gluon.}
    \label{feynfig:1}
    \end{center}
\end{figure}

\begin{figure}[t]
    \begin{center}
    \leavevmode
    \includegraphics[height=3cm,bb=106 629 529 725]{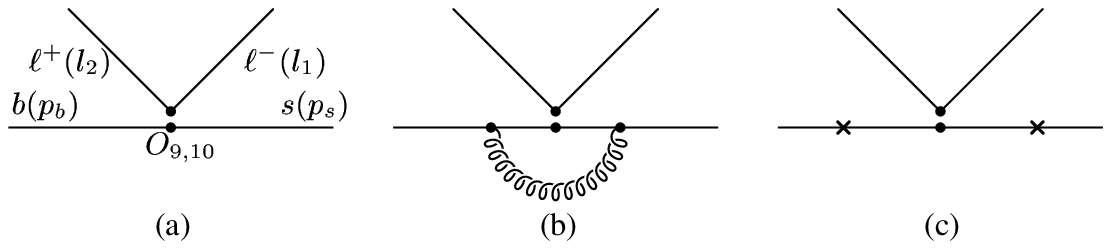}
    \vspace{2ex}
    \caption{Feynman diagrams associated with the operators $O_9$
    and $O_{10}$.
    (a) shows the lowest order diagrams, (b) and (c) show virtual- and
    bremsstrahlung corrections, respectively.}
    \label{feynfig:2}
    \end{center}
\end{figure}

As mentioned above, the virtual corrections to the matrix 
elements of the operators
$O_7$, $O_9$ and $O_{10}$, shown in figs. \ref{feynfig:1}b) and
\ref{feynfig:2}b),
suffer from infrared- and collinear singularities. According to the KNL
theorem, these singularities cancel when taking into account the corresponding
bremsstrahlung corrections shown in  
figs. \ref{feynfig:1}c) and \ref{feynfig:2}c). As these cancellations
only happen at the level of the decay rate, both virtual- and bremsstrahlung
corrections have to be regularized.
As in our previous works on $d\Gamma/d\sh$, we use for the
derivation of the double differential decay width a non-vanishing
strange quark mass as a regulator of the collinear singularities
and dimensional regularization ($d=4-2\epsilon$) 
for the infrared singularities. 
In the usual derivation of the decay width $d\Gamma/d\sh$
one integrates out the lepton variables in the first
step, after inserting a factor 1 in the form of eq. (\ref{deltainsertion}).
It turns out that in the Dirac trace of the lepton tensor $L_{\mu \nu}$ 
the terms with an odd number of $\gamma_5$ matrices become zero after
this integration. Furthermore, as the integrated lepton tensor
is symmetric in $\mu$ and $\nu$, it follows that also
in the hadron tensor (which is contracted with the lepton tensor
over the indices $\mu$ and $\nu$) only traces with an even number
of $\gamma_5$ matrices survive. 
Therefore, the $\gamma_5$ problems which usually appear in $d$-dimensions,
can be avoided when calculating $d\Gamma/d\sh$.
These statements are no longer true
if one aims to calculate the double differential decay width, which
means that traces with an odd number of $\gamma_5$ matrices are unavoidable.

In our derivation of the virtual corrections to the double differential
decays width, we calculated the loop corrections to the matrix elements
as in our previous papers \cite{Asatrian:2001de,Asatrian1}, viz.
using  anticommuting $\gamma_5$ and letting propagate all $d$ polarizations
of the virtual gluon in the loop. Using $(d-1)$-dimensional rotation
invariance, the momenta of the external particles
can be assumed to lie in four dimensions. Therefore, 
to proceed from the regulated matrix elements to the
double differential decay width, we do the remaining Dirac
algebra in $d=4$ dimensions. The subsequent phase space integrals
are, however, treated in $d$ dimensions.

We now turn to the bremsstrahlung corrections.
When calculating the squares of the matrix elements associated with
$O_7$, $O_9$, $O_{10}$ (and interference terms) some care has to be taken
in order to do the infrared regularization consistently.
As in the virtual corrections all $d$ gluon polarizations were allowed to
propagate, we have to emit all $d-2$ transverse polarizations
in the bremsstrahlung process.
As shown in refs. \cite{Greub89,GreubMink89}, this can be implemented
by doing the Dirac algebra in $d=4$ by summing the contributions 
{}from the emission of a gluon with
the 2 possible transverse directions in four dimensions 
(characterized by normal 4-dimensional polarization vectors),
and from the emission of the $(d-4)$ transverse polarizations
showing in the $d-4$ extra dimensions. Each of the latter couples
to the quarks (which remain in four dimensions) with a $\gamma_5$.
The subsequent phase space integrations are again worked out in 
$d$-dimensions.
\section{Phase space}
\label{sec:six}
\subsection{Fully differential phase space formula for lepton pair at rest}
Starting from the well-known expression for the differential decay width
for the process $b \to s \ell^+ \ell^-$ and inserting a unit factor
according to eq. (\ref{deltainsertion}), one obtains
\begin{equation}
\label{widthusual}
d \Gamma^{\text{b-rest}}(b \to s  \ell^+ \ell^- ) =
\frac{\overline{|M|^2}}{2 \, m_b} \, D\Phi^{\text{b-rest}} \, ,
\end{equation}
where $\overline{|M|^2}$ is the squared matrix element, summed
and averaged over spins and colors of the particles in
the final- and initial state, respectively. 
Note that in our application $\overline{|M|^2}$ depends only on scalar
products of four-vectors.
$D\Phi^{\text{b-rest}}$ is
the phase space factor which can be written as
\begin{eqnarray}
\label{phasefactors}
\nonumber
&&D\Phi^{\text{b-rest}}=D\Phi^{\text{b-rest}}_1 \, 
D\Phi^{\text{b-rest}}_2 \, ds \, , \\
&&D\Phi^{\text{b rest}}_1=(2\pi)^d \, \frac{d^{d-1}q}{2q^0} \, 
\frac{d^{d-1}p_s}{(2\pi)^{d-1}2p_s^0} \, \delta^d(p_b-p_s-q) \, , \\
\nonumber
&&D\Phi^{\text{b-rest}}_2=\frac{d^{d-1}l_1}{(2\pi)^{d-1}2l_1^0} \, 
\frac{d^{d-1}l_2}{(2\pi)^{d-1}2l_2^0} \, \delta^d(q-l_1-l_2) \, .
\end{eqnarray}
$p_b$, $p_s$, $l_1$, $l_2$ denote the
four-momenta of the $b$-quark, the $s$-quark, the negatively and positively
charged leptons, respectively, while $q=(l_1+l_2)$,
$q^0=\sqrt{\vec{q}^2+s}$ and
$s=q^2$. Note that eqs. (\ref{widthusual}) and (\ref{phasefactors})
generate the correct distributions of the decay products for a
$b$-quark decay at rest or with fixed velocity.

Our main goal is to calculate the double differential decay width 
$\frac{d^2\Gamma(b\to X_s l^+l^-)}{d\hat{s}dz}$, where $\hat{s}
=s/m_b^2$ and $z=\cos\theta$ with $\theta$ being 
the angle between the
momenta of the $b$-quark and the $\ell^+$, measured
in the rest frame  of the $(\ell^+\ell^-)$-pair. For this purpose
it is convenient to
first derive a fully differential phase space formula
in the rest frame of the lepton pair. 
In the following, unprimed momenta refer to the rest frame of the $b$-quark
and primed ones to the corresponding momenta in the rest frame of the
lepton pair. While in the rest frame
of the $b$-quark the value of the vector $\vec{q} = \vec{l}_1+\vec{l}_2$
varies from event to event, it is $\vec{p'_b}$ which varies from
event to event in the rest frame of the lepton pair.
%Starting from the usual phase space formula for a $b$-quark at
%rest, we substitue the variable $\vec{q}$ with $\vec{p'}_b$.
The relation between $\vec{q}$ and $\vec{p'}_b$
can be found from the equation
\begin{equation}
p'_b = \Lambda_q \, p_b \, ; \quad p_b=(m_b,\vec{0}) \, ,
\end{equation}
where $\Lambda_q$ is the Lorentz boost, which transforms the vector
$\vec{q}$ to rest. We obtain 
\begin{eqnarray}
\vec{p'_b}=-\frac{m_b}{\sqrt{s}} \vec{q} \quad 
(\mbox{and} \, \, \, p_b^{'0}=\frac{m_b}{\sqrt{s}} \, q^{0})\, .
\end{eqnarray} 
In the expression for the decay width this relation is most easily implemented
by multiplying eq. (\ref{widthusual}) with a factor 1 in the form
\begin{equation}
\label{adddelta}
1=\int \, d^{d-1} p'_b \, \delta^{d-1}(\vec{p'}_b + \frac{m_b}{\sqrt{s}} \,
\vec{q}) \, .
\end{equation}
We anticipate that the integration over the variable $\vec{q}$ will
finally perform the variable transformation $\vec{q} \leftrightarrow
\vec{p'}_b$. However, before doing this step we express all the
unprimed momenta in the matrix element squared and in the delta functions
with their primed counterparts, e.g. $l_2=\Lambda_q^{-1} l_2'$, etc..
Note that due to Lorentz invariance of 
$\overline{|M|^2}$, this quantity
is independent of $\Lambda_q^{-1}$, and therefore independent of $\vec{q}$.
The same is also true for the measure factors
of the final state particles and for the $d$-dimensional $\delta$-functions
in eq. (\ref{phasefactors}). 
The only remaining $\vec{q}$ dependence is contained in the term
\[
\frac{d^{d-1}q}{2\,q^0} \,
\delta^{d-1}(\vec{p'_b} + \frac{m_b}{\sqrt{s}} \,
\vec{q}) \, . \]
Integrating this eq. over $\vec{q}$, one obtains
\[
\int \frac{d^{d-1}q}{2\,q^0} \,
\delta^{d-1}(\vec{p'_b} + \frac{m_b}{\sqrt{s}} \,
\vec{q}) = 
\left( \frac{\sqrt{s}}{m_b} \right)^{d-2} \,
\frac{1}{2\,p_b^{'0}}  \, . 
\] 
To summarize:  The expression for the fully differential decay width
$d \Gamma(b \to s \ell^+ \ell^-)$ in the {\it rest frame of the
lepton pair} can be written as 
\begin{equation}
\label{doublewidthvirt}
d \Gamma(b \to s  \ell^+ \ell^- ) =
\frac{\overline{|M|^2}}{2 \, m_b} \, D \Phi \, ,
\end{equation}
with
\begin{eqnarray}
\label{phasefactorsleptrest}
\nonumber
&&D\Phi^{}=D\Phi^{}_1\, D\Phi_2 \, ds\, ,\\
&&D\Phi^{}_1=(2\pi)^d \, 
\left( \frac{\sqrt{s}}{m_b} \right)^{d-2} \,
\frac{d^{d-1} p_b}{2 p_b^0} \, 
\frac{d^{d-1}p_s}{(2\pi)^{d-1}2p_s^0} \, \delta^d(p_b-p_s-q)\, ,\\
\nonumber
&&D\Phi_2=\frac{d^{d-1}l_1}{(2\pi)^{d-1}2l_1^0} \, 
\frac{d^{d-1}l_2}{(2\pi)^{d-1}2l_2^0} \, \delta^d(q-l_1-l_2) \, .
\end{eqnarray}
As all momenta refer to the rest frame of the lepton pair, we omitted
the primes in eqs. (\ref{doublewidthvirt}) and (\ref{phasefactorsleptrest}).

For the case of real gluon emission, $b\to s g \ell^+ \ell^-$, 
the expression for the fully differential decay width
in the rest frame of the $(\ell^+ \ell^-)$-pair
can be derived in an analogous way. We obtain  
\begin{eqnarray}
\label{doublewidthbrems}
\nonumber
&&d \Gamma(b \to s g \ell^+ \ell^- ) =
\frac{\overline{|M|^2}}{2 \, m_b} \, D \Phi^{\text{brems}}\, ,  \\
&&D\Phi^{\text{brems}}=D\Phi^{\text{brems}}_1 \, D\Phi_2 \, ds \, ,\\
&&D\Phi^{\text{brems}}_1=(2\pi)^d \, 
\left( \frac{\sqrt{s}}{m_b} \right)^{d-2} \, 
\frac{d^{d-1}p_b}{2p_b^0} \,
\frac{d^{d-1}p_s}{(2\pi)^{d-1}2p_s^0} \,
\frac{d^{d-1}r}{(2\pi)^{d-1}2r^0} \, \delta^d(p_b-r- p_s-q) \, .
\end{eqnarray}
$D \Phi_2$ is the same as in eq. (\ref{phasefactorsleptrest}) and 
$r$ is the four-momentum of the gluon. 
\subsection{Phase space integrations}
In this subsection we present the results for the phase space
formulas for the double differential decay width
where we integrate over the variables constrained by 
the $\delta$-functions and over the variables on which 
$\overline{|M|^2}$ does not depend.

To get the desired expression for the bremsstrahlung
process, we start from eq. (\ref{doublewidthbrems})
and integrate over $\vec{l}_1$ and $\vec{p}_s$ by making use
of the spacial parts of the two $d$-dimensional $\delta$-functions.
Using then rotation
invariance in ($d-1$) dimensions, we can assume that
in $\overline{|M|^2}$ the ``three-momenta''
of the remaining particles have the form
\bea
\label{vectorform}
\vec{p}_b &=& (|\vec{p}_b|,0,0;....) \, , \nonumber \\
\vec{l}_2 &=& (E_2 \cos \theta, E_2 \sin \theta,0;....) \, , \nonumber \\
\vec{r} &=& (E_r \cos \theta_1, E_r \sin \theta_1 \cos \theta_2,
                              E_r \sin \theta_1 \sin \theta_2;....) \, ,
\eea
where the dots symbolize the components of
extra space dimensions, which are all zero. $E_2$ and $E_r$
are the energies of the massless positively charged lepton and
the gluon, respectively. Making use of the remaining two
one-dimensional  $\delta$-functions, 
we can express $E_2$ and $\theta_1$ in terms of the other variables
as
\begin{equation}
E_2 = \frac{\sqrt{s}}{2} \quad ; \quad
\cos \theta_1 = \frac{2 E_b \sqrt{s} - 2 E_r \sqrt{s}  + 2 E_r E_b -s - m_b^2
  + m_s^2}{2 E_r |\vec{p}_b|} \, .
\end{equation}
$E_b$ is the energy of the $b$-quark and $\sqrt{s}$ is the invariant mass of 
the lepton pair. After integration over the additional polar angles
of $\vec{p}_b$, $\vec{l}_2$ and $\vec{r}$, on which $\overline{|M|^2}$
does not depend, we obtain
(using $z=\cos \theta$, $z_2=\cos \theta_2$,
$\sh=s/m_b^2=q^2/m_b^2$, $d=4-2\epsilon$)
\bea
\frac{d^2\Gamma(b \to s g \ell^+ \ell^-)}{d\sh \, dz } &=& 
\left( \frac{\mu^2 \, \mbox{e}^{\gamma}}{4\pi} \right)^{3\epsilon} \, 
\frac{m_b^{1-2\epsilon} \, 
\sh^{1-2\epsilon}}{(2\pi)^{3d-4} \, 2^{7-4\epsilon}} \,
\Omega_{d-1} \, \Omega_{d-2} \, \Omega_{d-3} \, \times \nonumber \\
&& 
\int \, \overline{|M|^2} \, 
W^{-\epsilon} \, (1-z^2)^{-\epsilon} \, (1-z_2^2)^{-1/2-\epsilon} \,
dz_2 \,
dE_r \, dE_b \, ,
\eea 
where $W$ reads
\[
W=4 \, (E_r-E_r^{\rm{min}}) \, (E_r^{\rm{max}}-E_r) \, (m_b^2 - 2 \sqrt{s} \, 
E_b +s) \, .
\]
The three factors $\Omega$ stem from the integration over the polar
angles on which  $\overline{|M|^2}$ does not depend (explicitly,
$\Omega_d=2\pi^{d/2}/\Gamma(d/2)$).
The boundaries of the integration variables are
\bea
E_r^{\rm{min}}=\frac{m_b^2+s-2E_b\sqrt{s}-m_s^2}{2(E_b+|\vec{p}_b|-\sqrt{s})}
\le & E_r & \le \frac{m_b^2+s-2E_b\sqrt{s}-m_s^2}{2(E_b-|\vec{p}_b|-\sqrt{s})}=
E_r^{\rm{max}} \, ,\nonumber \\
m_b \le & E_b & \le \frac{m_b^2+s-m_s^2}{2\sqrt{s}} \, , \nonumber \\
-1 \le &z_2& \le 1 \, .
\eea 

To get the corresponding expression for the double differential decay width for
the process $b \to s \ell^+ \ell^-$, we start from eq. (\ref{doublewidthvirt})
and integrate over $\vec{l}_1$ and $\vec{p}_s$ by making use of the
spacial parts of the two $d$-dimensional $\delta$-functions.
Using rotation invariance, the three momenta of the remaining particles
($b$-quark and $\ell^+$) can be assumed to have the form as in eq. 
(\ref{vectorform}). 
The remaining two 
one-dimensional $\delta$-functions can be used to express 
the energy $E_b$ of the  $b$-quark and the energy  $E_2$ of $\ell^+$
in terms of $s$. Explicitly, we obtain
\[ E_b = \frac{m_b^2+s-m_s^2}{2\sqrt{s}} \quad ; \quad
   E_2 = \frac{\sqrt{s}}{2} \, .
\] 
After integration over the  angles
of $\vec{p}_b$ and $\vec{l}_2$, on which $\overline{|M|^2}$
does not depend, we obtain
\bea
\frac{d^2\Gamma(b \to s  \ell^+ \ell^-)}{d\sh \, dz } &=& 
\left( \frac{\mu^2 \, \mbox{e}^{\gamma}}{4\pi} \right)^{2\epsilon} \, 
\frac{m_b^{-2\epsilon} \, 
\sh^{1/2-2\epsilon}}{(2\pi)^{2d-3} \, 2^{6-2\epsilon}} \,
\Omega_{d-1} \, \Omega_{d-2}  \, 
\overline{|M|^2} \, 
 (1-z^2)^{-\epsilon} \, |\vec{p}_b|^{d-3} \, .
\eea 
\section{Calculation of the sum of virtual- and bremsstrahlung
corrections associated with $\boldsymbol{O_7}$, $\boldsymbol{O_9}$ and 
$\boldsymbol{O_{10}}$}
\label{sec:seven}
In this section, we explain in some detail
the tricks which allow to construct the functions
$f_{99}$ and $f_{910}$ in eq. (\ref{doublewidth}) in a simplified manner. 
The other functions
$f_{77}$, $f_{79}$ and $f_{710}$ can be obtained in an analogous way.
We use the notations
\[
\Gamma_{ij}(\sh,z) = \frac{d^2\Gamma_{ij}}{d\sh \, dz} \quad
\mbox{and} \quad
\Gamma_{ij}(\sh) = \frac{d\Gamma_{ij}}{d\sh} \quad (i \le j)
\]   
for the contributions of the pair $(\wtO_i,\wtO_j)$ to the double
differential decay width and to the invariant mass distribution,
respectively. To make explicit the lowest order piece ($^0$), 
the virtual- ($^v$) and bremsstrahlung ($^b$) corrections, we 
write
\bea
\Gamma_{ij}(\sh,z) &=& \Gamma_{ij}^0(\sh,z) + 
\Gamma_{ij}^v(\sh,z) + 
\Gamma_{ij}^b(\sh,z) \, ,\nonumber \\ 
\Gamma_{ij}(\sh) &=& \Gamma_{ij}^0(\sh) + 
\Gamma_{ij}^v(\sh) + 
\Gamma_{ij}^b(\sh) \, . 
\eea
As mentioned in section \ref{sec:three}, the virtual corrections to the
matrix element were written in our earlier papers as multiples of
tree-level matrix elements, explicitly
\bea
\bra s \ell^+ \ell^- | \wtO_7 |b \ket_{\rm{virt}}  &=& 
-\frac{\alpha_s}{4\pi} \,
F_7^{(7)} \, \bra \wtO_7 \ket_{\rm{tree}} 
-\frac{\alpha_s}{4\pi} \,
F_7^{(9)} \, \bra \wtO_9 \ket_{\rm{tree}} \, ,\nonumber \\
\bra s \ell^+ \ell^- | \wtO_9 |b \ket_{\rm{virt}}  &=& 
-\frac{\alpha_s}{4\pi} \,
F_9^{(7)} \, \bra \wtO_7 \ket_{\rm{tree}} 
-\frac{\alpha_s}{4\pi} \,
F_9^{(9)} \, \bra \wtO_9 \ket_{\rm{tree}} \, ,\nonumber \\
\bra s \ell^+ \ell^- | \wtO_{10} |b \ket_{\rm{virt}}  &=& 
-\frac{\alpha_s}{4\pi} \,
F_9^{(7)} \, \bra \wtO_{7,5} \ket_{\rm{tree}} 
-\frac{\alpha_s}{4\pi} \,
F_9^{(9)} \, \bra \wtO_{10} \ket_{\rm{tree}} \, .
\eea
Note that $\bra \wtO_{7,5}\ket_{\rm{tree}}$ is obtained from 
$\bra \wtO_{7}\ket_{\rm{tree}}$ by replacing the lepton vector current
by the corresponding axial vector current.
As the explicit form of the (infrared singular) functions 
$F_9^{(9)}$ and $F_7^{(7)}$
is not needed in the following construction, 
we only list $F_9^{(7)}$ and $F_7^{(9)}$:
\begin{equation}
F_9^{(7)}(\sh) = \frac{2}{3} \ln (1-\sh) \, ; \quad  
F_7^{(9)}(\sh) = \frac{16}{3 \, \sh} \ln(1-\sh) \, .
\end{equation}
\subsection{Construction of $\boldsymbol{f_{99}(\sh,z)}$ }
The virtual corrections to the double- or single differential
decay width are now readily obtained. 
For $\Gamma_{99}^v(\sh,z)$ and $\Gamma_{99}^v(\sh)$ we get
\bea
\Gamma_{99}^v(\sh,z) &=& 
-\frac{2 \alpha_s}{4\pi} F_9^{(9)}(\sh) \, \Gamma^0_{99}(\sh,z) 
-\frac{\alpha_s}{4\pi} F_9^{(7)}(\sh) \, \Gamma^0_{79}(\sh,z) \, ,
\nonumber \\
\Gamma_{99}^v(\sh) &=& 
-\frac{2 \alpha_s}{4\pi} F_9^{(9)}(\sh) \, \Gamma^0_{99}(\sh) 
-\frac{\alpha_s}{4\pi} F_9^{(7)}(\sh) \, \Gamma^0_{79}(\sh) \, .
\eea
We note that $\Gamma^0_{99}(\sh,z)$ and $\Gamma^0_{99}(\sh)$
are understood to be
evaluated in $d$-dimensions as described in sections
\ref{sec:five} and \ref{sec:six}, 
because the function $F_9^{(9)}$ is infrared singular.

\noindent
We now turn to the crucial point of our construction,
which drastically simplifies the calculation of the bremsstrahlung
corrections. We form the combination  
\begin{equation}
\label{combvirt}
\hat{\Gamma}^v_{99}(\sh,z) = \Gamma^v_{99}(\sh,z) -
\frac{\Gamma^0_{99}(\sh,z)}{\Gamma^0_{99}(\sh)} \, \Gamma^v_{99}(\sh) \, ,
\end{equation}
in which the contributions proportional to the singular function
$F_9^{(9)}$ drop out completely. $\hat{\Gamma}^v_{99}(\sh,z)$ 
is therefore finite. Explicitly, we get
\begin{equation}
\label{combvirtexpl}
\hat{\Gamma}^v_{99}(\sh,z)=\frac{\alpha_s}{4\pi} \,
F_9^{(7)}(\sh) \, \frac{m_b^5 \, \alpha_{\text{em}}^2 \, G_F^2 \, 
|V_{tb} V_{ts}^*|^2 \, \wtC_9^2 \, 
(1-\sh)^3 \,(1-3z^2)}{256 \pi^5 \, (1+2\sh)}.
\end{equation}
 
\noindent
We now form the analogous combination for the bremsstrahlung
corrections, viz.
\begin{equation}
\label{combbrems}
\hat{\Gamma}^b_{99}(\sh,z) = \Gamma^b_{99}(\sh,z) -
\frac{\Gamma^0_{99}(\sh,z)}{\Gamma^0_{99}(\sh)} \, \Gamma^b_{99}(\sh) \, .
\end{equation}
It follows from the Kinoshita-Lee-Neuenberg (KLN) theorem that
$\hat{\Gamma}^b_{99}(\sh,z)$ must also be finite.
Using eqs. (\ref{combvirt}) and (\ref{combbrems}), one can write
the sum of the virtual- and bremsstrahlung corrections to the double
differential decay width in the form
\begin{equation}
\label{combsum}
 \Gamma^v_{99}(\sh,z) + \Gamma^b_{99}(\sh,z)  = 
\hat{\Gamma}^v_{99}(\sh,z)+
\hat{\Gamma}^b_{99}(\sh,z) +
\frac{\Gamma^0_{99}(\sh,z)}{\Gamma^0_{99}(\sh)} \, 
\left( \Gamma^v_{99}(\sh) + \Gamma^b_{99}(\sh) \right)    \, .
\end{equation}
$\hat{\Gamma}^v_{99}(\sh,z)$ on the r.h.s of eq. (\ref{combsum}) is given
in eq. (\ref{combvirtexpl}). $\left( \Gamma^v_{99}(\sh) + \Gamma^b_{99}(\sh)
\right)$ is also known, viz.
\begin{equation}
\label{w99comb}
\Gamma^v_{99}(\sh) + \Gamma^b_{99}(\sh) = \Gamma^0_{99}(\sh) \left(
1 + \frac{2 \, \alpha_s}{\pi} \, \omega_{99}(\sh) \right) \, ,
\end{equation}
where $\omega_{99}(\sh)$ is given in refs.
\cite{Asatrian:2001de,Asatrian1} (see also eq. (\ref{omega99})). 
$\Gamma^0_{99}(\sh,z)$ which in eq. (\ref{combsum})
is only needed in $d=4$ dimensions,
reads
\begin{equation}
\label{gamma990}
\Gamma^0_{99}(\sh,z)=\frac{m_b^5 \, \alpha_{\text{em}}^2 \, G_F^2 \, 
|V_{tb} V_{ts}^*|^2 \, \wtC_9^2 }{1024 \pi^5} \, 
(1-\sh)^2  \left[(1-z^2)+\sh (1+z^2) \right] \, .
\end{equation}
This implies that the sum of virtual- and bremsstrahlung corrections
to the double differential decay width, and hence the function $f_{99}(\sh,z)$
in eq. (\ref{rarewidth}),  
is easily obtained once the
finite combination $\hat{\Gamma}^b_{99}(\sh,z)$ in eq. (\ref{combbrems}) is known.
\subsection{Construction of $\boldsymbol{f_{910}(\sh)}$ }
As $\Gamma_{910}^0(\sh)$ turns out to be zero, one cannot take
the combination analogous to eq. (\ref{combvirt}). Instead, we use
the combination  
\begin{equation}
\label{combvirta}
\hat{\Gamma}^v_{910}(\sh,z) = \Gamma^v_{910}(\sh,z) -
\frac{\Gamma^0_{910}(\sh,z)}{\Gamma^0_{99}(\sh)} \, \Gamma^v_{99}(\sh) \, .
\end{equation}
Again, the part
proportional to the singular function
$F_9^{(9)}$ drops out and  $\hat{\Gamma}^v_{910}(\sh,z)$ 
is finite. Explicitly, we find
\begin{equation}
\label{combvirtaexpl}
\hat{\Gamma}^v_{910}(\sh,z)=\frac{2\alpha_s}{4\pi} \,
F_9^{(7)}(\sh) \, \frac{m_b^5 \, \alpha_{\text{em}}^2 \, G_F^2 \, 
|V_{tb} V_{ts}^*|^2 \, \wtC_9 \, \wtC_{10} \, 
(1-\sh)^3 \, z}{128 \pi^5 \, (1+2\sh)} .
\end{equation}
The  analogous combination for the bremsstrahlung
corrections, viz.
\begin{equation}
\label{combbremsa}
\hat{\Gamma}^b_{910}(\sh,z) = \Gamma^b_{910}(\sh,z) -
\frac{\Gamma^0_{910}(\sh,z)}{\Gamma^0_{99}(\sh)} \, \Gamma^b_{99}(\sh) \, 
\end{equation}
is also finite. {}From
eqs. (\ref{combvirta}) and (\ref{combbremsa}) one gets
\begin{equation}
\label{combsuma}
 \Gamma^v_{910}(\sh,z) + \Gamma^b_{910}(\sh,z)  = 
\hat{\Gamma}^v_{910}(\sh,z)+
\hat{\Gamma}^b_{910}(\sh,z) +
\frac{\Gamma^0_{910}(\sh,z)}{\Gamma^0_{99}(\sh)} \, 
\left( \Gamma^v_{99}(\sh) + \Gamma^b_{99}(\sh) \right)    \, .
\end{equation}
$\hat{\Gamma}^v_{910}(\sh,z)$  and  $\left( \Gamma^v_{99}(\sh) +
  \Gamma^b_{99}(\sh) \right)$ are given  in eqs. (\ref{combvirtaexpl})
and (\ref{w99comb}), respectively. $\Gamma^0_{910}(\sh,z)$ 
which is only needed
in $d=4$ dimensions in eq. (\ref{combsuma}), reads 
\begin{equation}
\Gamma^0_{910}(\sh,z)=-\frac{m_b^5 \, \alpha_{\text{em}}^2 \, G_F^2 \, 
|V_{tb} V_{ts}^*|^2 \, \wtC_9 \, \wtC_{10} }{256 \pi^5} \, 
(1-\sh)^2  \, \sh \, z \, .
\end{equation}
This implies that the function $f_{910}(\sh)$  
is easily obtained, once the
finite combination $\hat{\Gamma}^b_{910}(\sh,z)$ 
defined in eq. (\ref{combbremsa}) is known.

To obtain the functions 
$f_{77}(\sh,z)$, $f_{79}(\sh,z)$ and $f_{710}(\sh)$,
one can proceed in a similar way. Forming suitable combinations,
the hardest part of the calculation of these functions boils 
down to the evaluation of a finite combination of bremsstrahlung terms.

A remark concerning to the evaluation of the finite bremsstrahlung
combinations is in order: We carefully investigated all five combinations
(needed to construct the five $f$-functions) in $d=4-2\epsilon$ dimensions,
as in principle terms of order $\epsilon^1$ from the phase space factors
could
multiply divergent integrals and in this way generate finite terms.
We found, however, that this case does not occur in our actual
calculations:  Expanding all
combinations up to order $\epsilon$ (or even $\epsilon^2$) before
doing the phase space integrations  
over the variables $E_r$ and $E_b$ (see section \ref{sec:six}), we
found that all the
occurring integrals are finite. This means, that it is correct to evaluate
the finite combinations in $d=4$ dimensions. 
\section{Phenomenological analysis}
\label{sec:eight}
\begin{figure}[htb]
\begin{center}
    \includegraphics[width=7.5cm, bb=19 179 581 580]{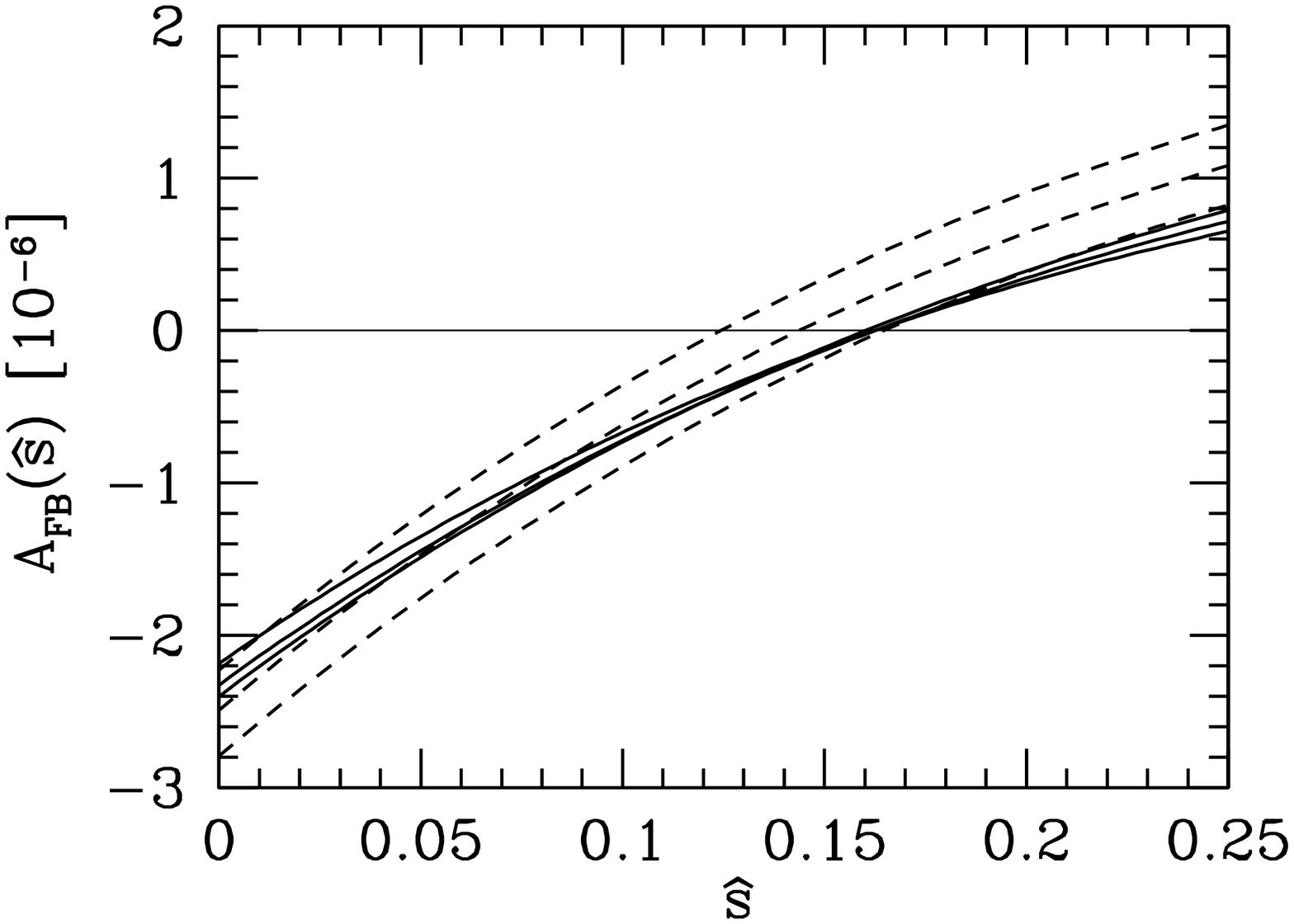}
    \includegraphics[width=7.5cm, bb=19 179 581 580]{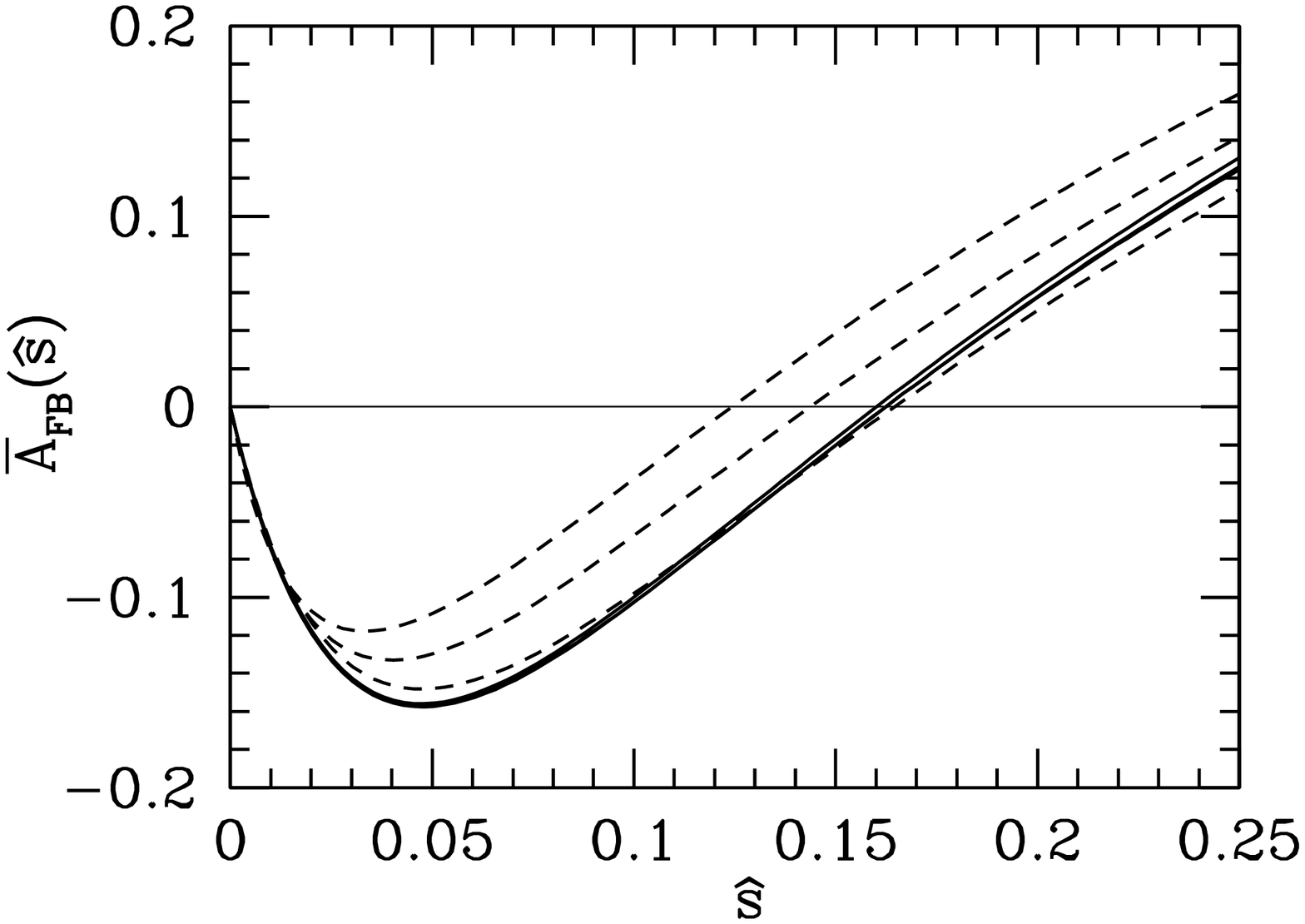}
    \vspace{0.5cm}
    \caption[]{Left frame: 
    Unnormalized forward-backward asymmetry $A_{\text{FB}}(\sh)$.
    The three solid lines
    show the NNLL prediction for $\mu=2.5,5.0,10.0$ GeV, respectively. 
    The corresponding curves in NLL approximation are shown by dashed lines.
    Right frame: 
    Normalized forward-backward asymmetry $\overline{A}_{\text{FB}}(\sh)$. The 
    lines have the same meaning as in the left frame. $m_c/m_b=0.29$.}
    \label{fig:1}
    \end{center}
\end{figure}
In this section, we mainly investigate the impact of the NNLL QCD
corrections on the forward-backward asymmetries defined in eqs.
(\ref{asymmnorm}) and (\ref{asymmunnorm}) in the standard model.
We restrict ourselves to the range of $\sh=s/m_b^2$ below 0.25, i.e., to
the region below the $J/\psi$ threshold. As our main emphasis is
to investigate the improvements in the perturbative part, in particular
the reduction of the renormalization scale dependence, we do not
include non-perturbative corrections, although in this $\sh$-region 
they are known to a large extent 
\cite{Falk:1994dh,Ali:1997bm,chen:1997,Buchalla:1998ky,Buchalla:1998mt,Krueger:1996}.
In our analysis, we use the following fixed
values for the input parameters: $m_b^{\rm{pole}}=4.8$ GeV, 
$\alpha_{\text{em}}=1/133$, $\mbox{BR}_{\text{sl}}=0.104$,
$m_t^{\rm{pole}}=174$ GeV, $\alpha_s(m_Z)=0.119$ and $|V_{tb} \, 
V_{ts}|/|V_{cb}|=0.976$. The values of $m_c/m_b$ and of the renormalization
scale $\mu$ are specified in the captions of the individual figures.

In figs. \ref{fig:1} we illustrate the reduction of the renormalization
scale dependence of the forward-backward asymmetries when
taking into account NNLL QCD corrections. As usual, the renormalization
scale is varied between 2.5 GeV and 10.0 GeV. For definiteness,
we should mention that in the unnormalized forward-backward asymmetry
$A_{\text{FB}}(\sh)$, 
we evaluated the denominator $\Gamma(B \to X_c e 
\bar{\nu}_e)$ in eq. (\ref{asymmunnorm}) always at $\mu=5$ 
GeV\footnote{We checked that the results only marginally change when
varying the scale also in the semileptonic decay width.}. 
The results are remarkable:
While the NLL asymmetries (shown by dashed lines for
$\mu$=2.5, 5.0 and 10 GeV) suffered from a relatively large renormalization
scale dependence, the theoretical uncertainty related to the 
choice of the renormalization scale is significantly reduced
at the NNLL level. For example, at $\sh=0$ we find
\begin{equation}
A_{\text{FB}}^{\text{NLL}}(0) = -(2.51 \pm 0.28) \times 10^{-6} \, ; \quad 
A_{\text{FB}}^{\text{NNLL}}(0) = -(2.30 \pm 0.10) \times 10^{-6} \, . 
\end{equation}
This corresponds to a reduction of the $\mu$-dependence from $\pm 11\%$
to $\pm 4.5\%$, which is similar to the situation found for the
differential branching ratio in ref. \cite{Asatrian1}.
When looking at the position
$\sh_0$, where the forward-backward asymmetries are zero, 
the reduction of the $\mu$-dependence at NNLL is even stronger.
We find (when only taking into
account the error due to the $\mu$-dependence) 
\begin{equation}
\label{sh0mu}
\sh_0^{\rm{NLL}} = 0.144 \pm 0.020 \, ; \quad
\sh_0^{\rm{NNLL}} = 0.162 \pm 0.002 \, .
\end{equation}
\begin{figure}[htb]
\begin{center}
    \includegraphics[width=7.5cm, bb=19 179 581 580]{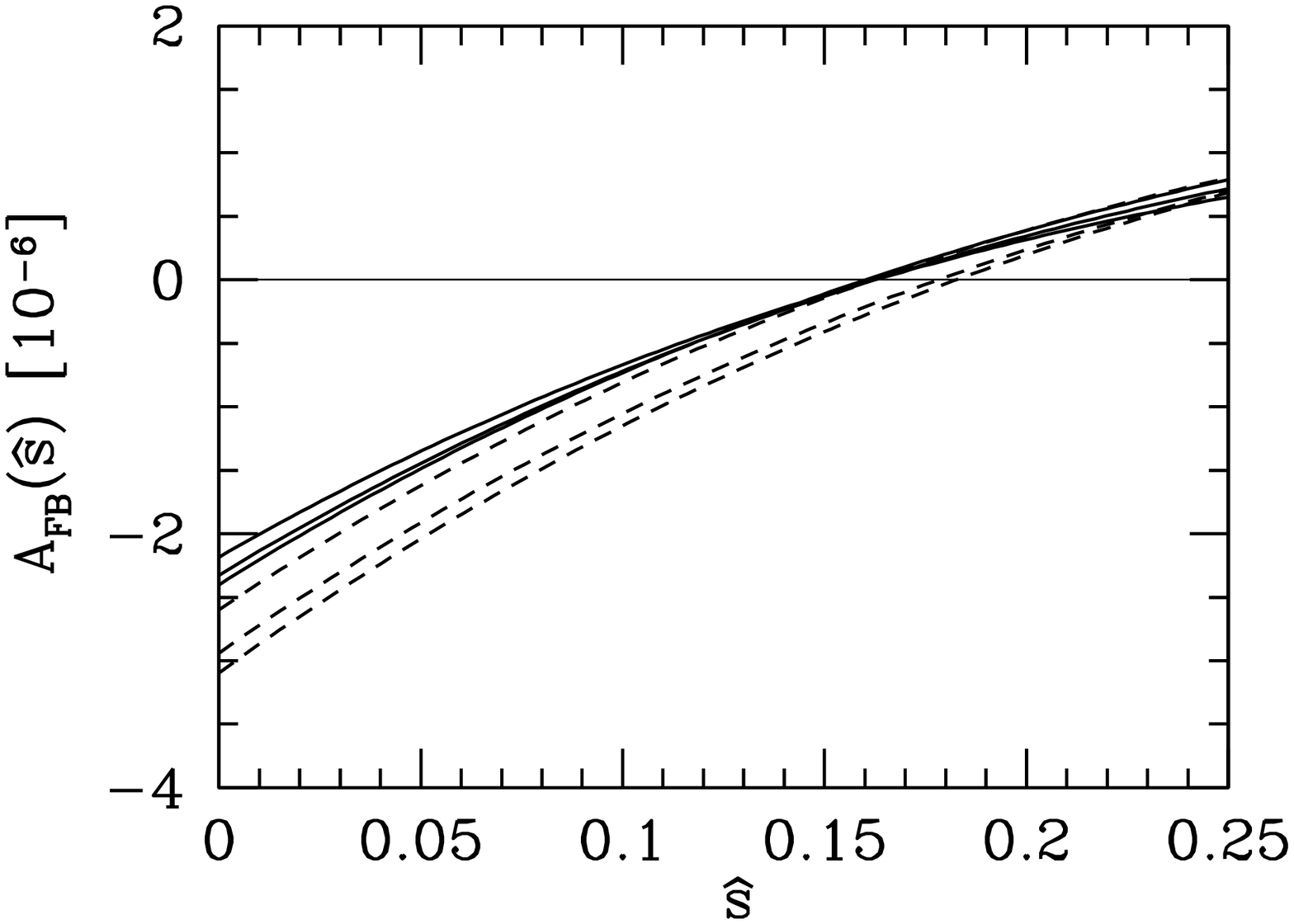}
    \includegraphics[width=7.5cm, bb=19 179 581 580]{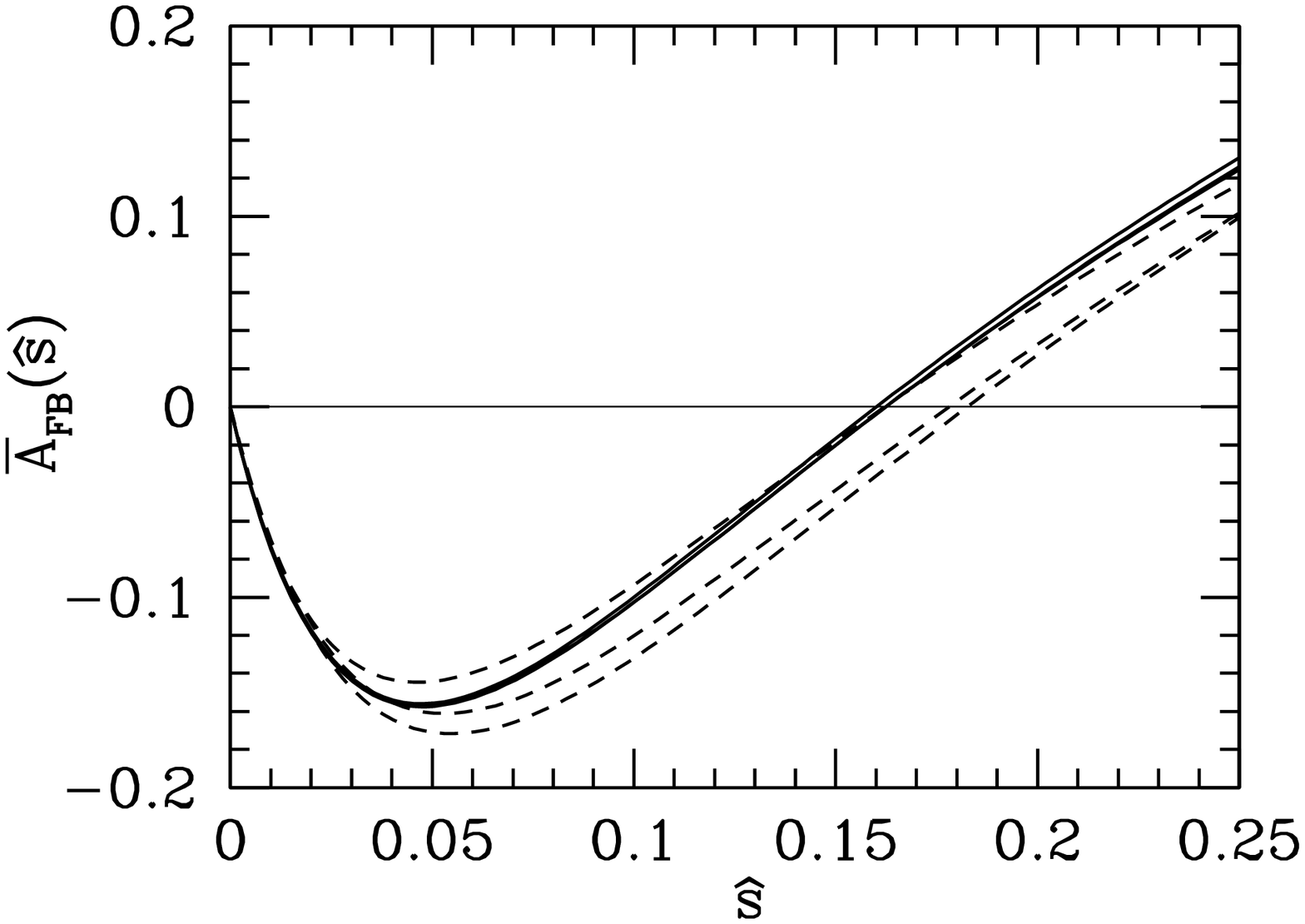}
    \vspace{0.5cm}
    \caption[]{Left frame: Unnormalized forward-backward asymmetry 
    $A_{\text{FB}}(\sh)$. 
    The three solid lines
    show the NNLL prediction for $\mu=2.5,5.0,10.0$ GeV, respectively. 
    The dashed lines show the corresponding results when switching
    off the functions $f_{710}(\sh)$ and $f_{910}(\sh)$. 
    Right frame: Normalized forward-backward asymmetry 
     $\overline{A}_{\text{FB}}(\sh)$. 
    The three solid lines
    show the NNLL prediction for $\mu=2.5,5.0,10.0$ GeV, respectively. 
    The dashed lines show the corresponding results when switching
    off the functions $f_{710}(\sh)$, $f_{910}(\sh)$, 
    $\omega_{77}(\sh)$, $\omega_{99}(\sh)$, and $\omega_{79}(\sh)$.
    $m_c/m_b=0.29$.}
    \label{fig:2}
    \end{center}
\end{figure}
The parts of the NNLL corrections to the forward-backward asymmetries 
which are contained in the effective Wilson coefficients
$\widetilde{C}_7^{\eff}$, 
$\widetilde{C}_9^{\eff}$ and
$\widetilde{C}_{10}^{\eff}$ (see eqs. (\ref{effcoeff9})-(\ref{effcoeff10})), 
i.e., the virtual corrections to the matrix elements 
of the operators $O_1$, $O_2$
and $O_8$ and the NNLL contributions to the Wilson coefficients,
are known for quite some time. In figs. \ref{fig:2}
we illustrate the importance
of the new contributions related to virtual- and bremsstrahlung corrections
to $O_7$, $O_9$ and $O_{10}$, which are encoded through the functions
$f_{710}(\sh)$ and $f_{910}(\sh)$. The solid lines show the full NNLL results,
while the dashed ones are obtained by switching off the functions 
$f_{710}(\sh)$ and $f_{910}(\sh)$ (in the case of the normalized
forward-backward asymmetry also the functions 
$\omega_{99}(\sh)$, $\omega_{77}(\sh)$ and $\omega_{79}(\sh)$ are switched
off).
We find that the new contributions
are crucial, in particular for
the reduction of the renormalization scale dependence.

\begin{figure}[htb]
\begin{center}
    \includegraphics[width=7.5cm, bb=19 179 581 580]{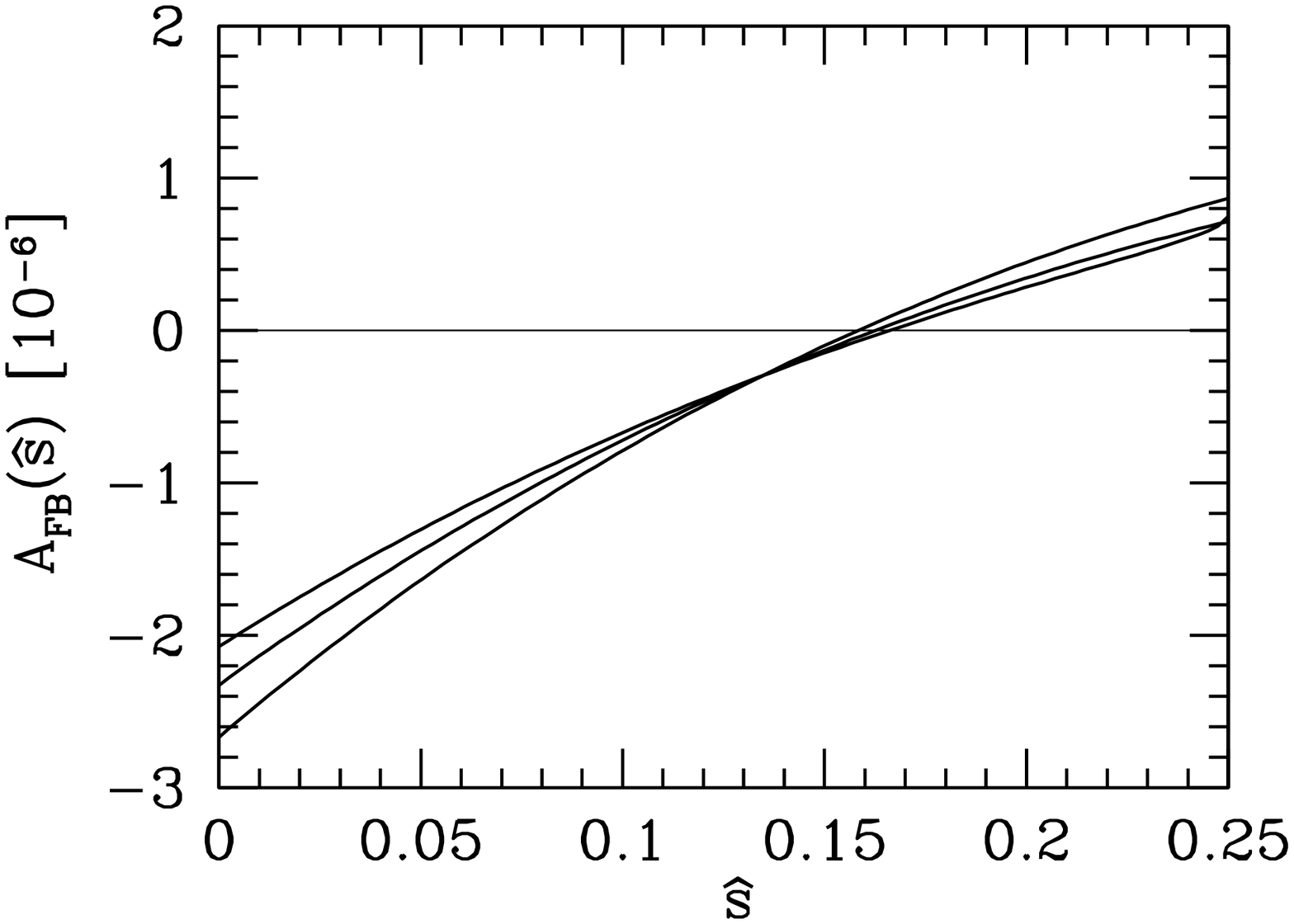}
    \includegraphics[width=7.5cm, bb=19 179 581 580]{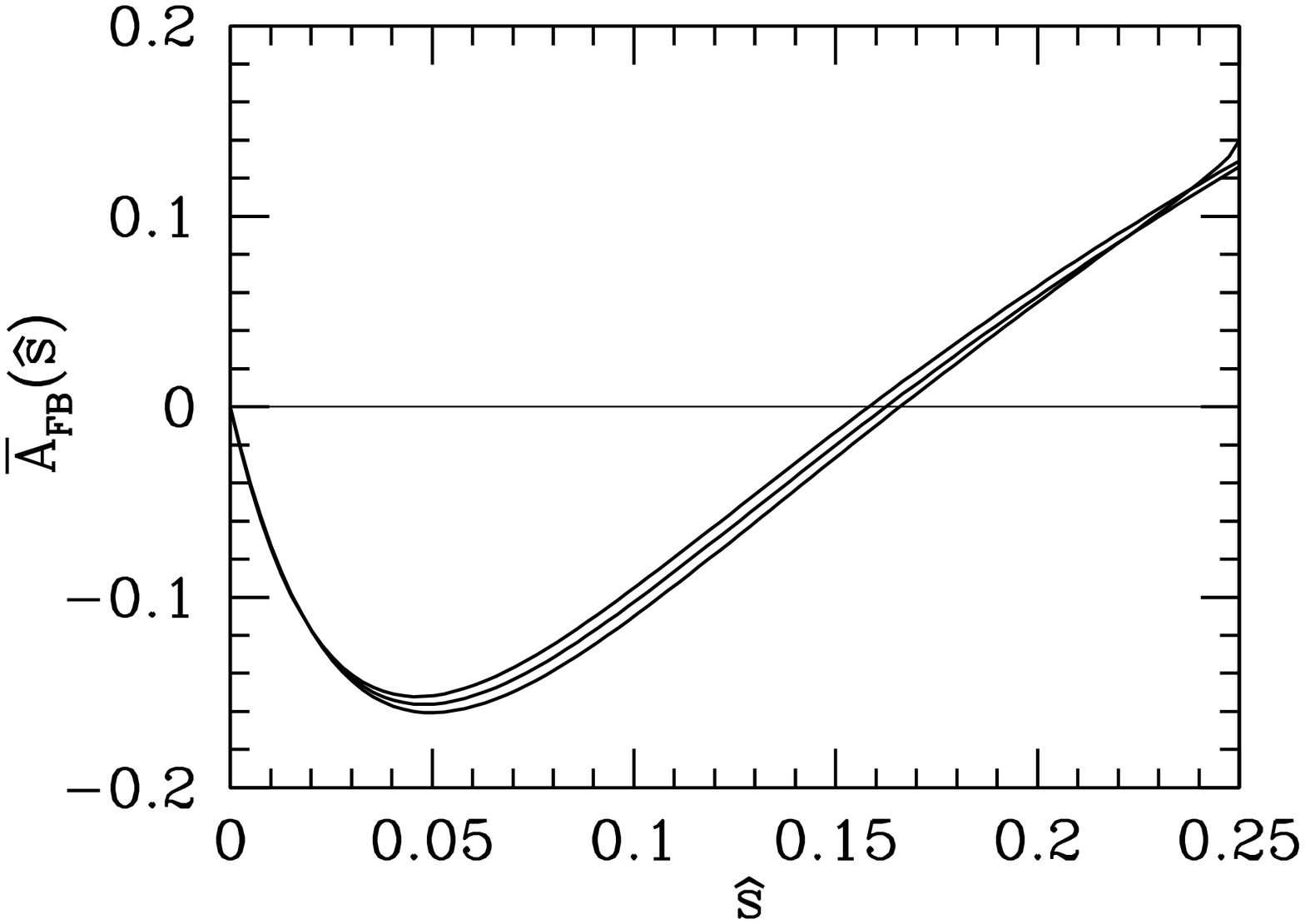}
    \vspace{0.5cm}
    \caption[]{Left frame: Unnormalized forward-backward asymmetry 
    $A_{\text{FB}}(\sh)$.
    The three lines show the NNLL prediction for 
    $m_c^{\pole}/m_b^{\pole} = 0.25, 0.29, 0.33 $, respectively.
    The renormalization scale is $\mu=5$ GeV. 
    Right frame: The same for the normalized forward-backward asymmetry 
    $\overline{A}_{\text{FB}}(\sh)$.}
    \label{fig:3}
    \end{center}
\end{figure}
As found in refs. \cite{Asatrian1,Asatrian2}, the error on the
decay width $d\Gamma(b \to X_s \ell^+ \ell^-)/d\sh$ due to uncertainties
in the input parameters is by far dominated by the uncertainty
of the charm quark mass $m_c$. 
In principle, there are two sources for this  uncertainty.
First, it is unclear whether $m_c$ in the virtual- and bremsstrahlung
corrections should be interpreted as the pole mass or 
the $\overline{\mbox{MS}}$ mass (at an appropriate scale).
Second, the question arises what the numerical value of $m_c$ is, 
once a choice concerning the definition of $m_c$ has
been made. These issues were investigated in detail in ref. \cite{Asatrian2}
and led to the conclusion that the error due to uncertainties in the
parameter $m_c/m_b$ is conservatively estimated when using for this quantity
$m_c^{\rm{pole}}/m_b^{\rm{pole}} = 0.29 \pm 0.04$. For a discussion
of the corresponding questions for the process $B \to X_s \gamma$, we refer to
\cite{Gambino01}. Motivated by these studies, we illustrate in figs. 
\ref{fig:3}
the dependence of the forward-backward asymmetries on 
$m_c^{\rm{pole}}/m_b^{\rm{pole}}$. The three lines
show the asymmetries for the values 
$m_c^{\rm{pole}}/m_b^{\rm{pole}}$=0.25, 0.29 and 0.33. 
We find that for most values of $\sh$ the charm quark mass dependence of 
the normalized
forward-backward asymmetry $\overline{A}_{\text{FB}}(\sh)$ is smaller than
the one of the unnormalized counterpart  $A_{\text{FB}}(\sh)$.
This is related to the fact
that a relatively large charm quark mass dependence enters the observable
$A_{\text{FB}}(\sh)$ through the semileptonic decay width present
in the defining eq. (\ref{asymmunnorm}); this is not the case
for the normalized version (see eq. (\ref{asymmnorm})).
For $\sh_0$, the position where the forward-backward asymmetries
vanish, we find (when taking into account only the error due to $m_c/m_b$)
\begin{equation}
\label{sh0mc}
\sh_0^{\rm{NNLL}} = 0.162 \pm 0.005 \, .
\end{equation}

%---------- figure ----------
\begin{figure}[htb]
\begin{center}
    \includegraphics[width=7.5cm, bb=19 179 581 580]{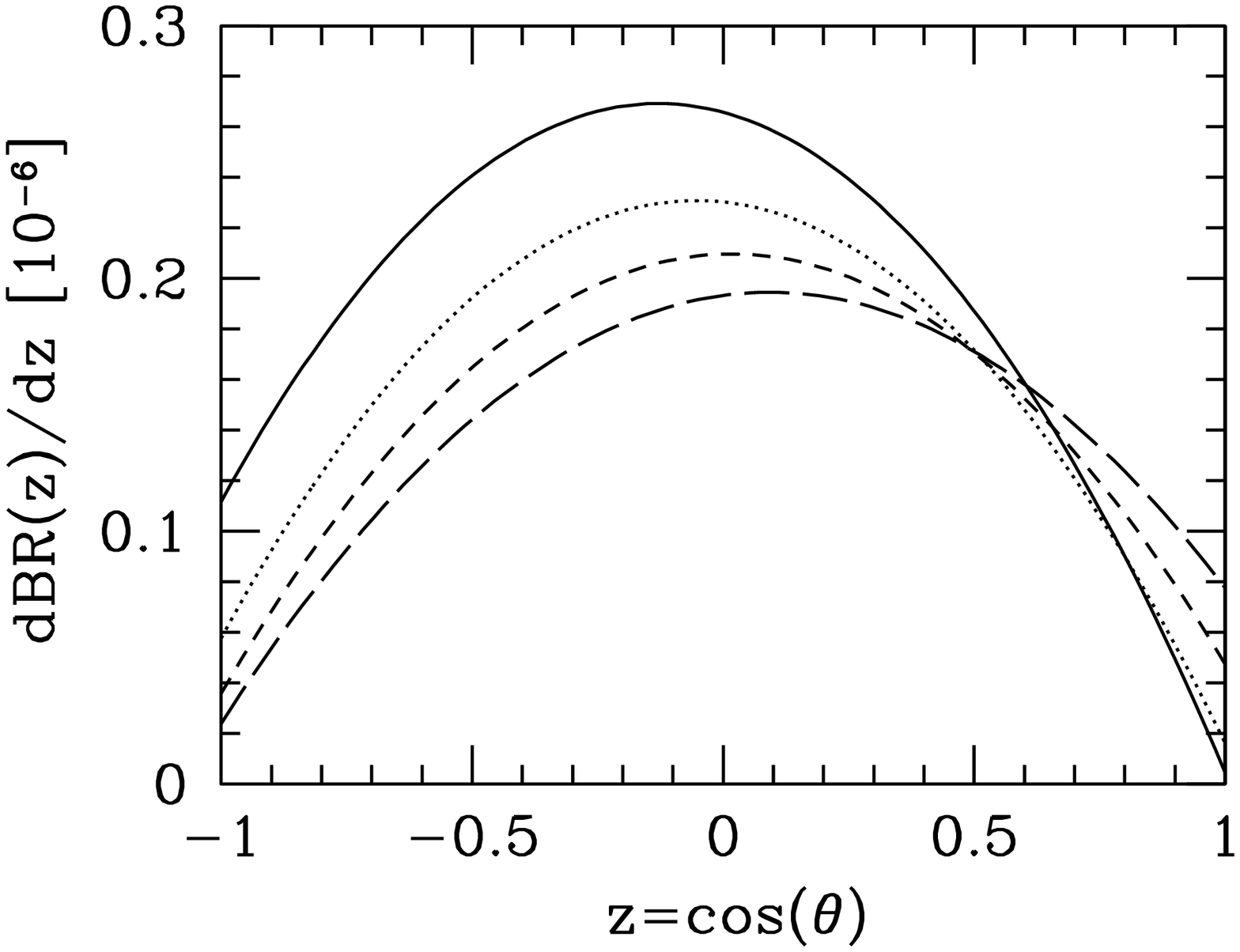}
    \includegraphics[width=7.5cm, bb=19 179 581 580]{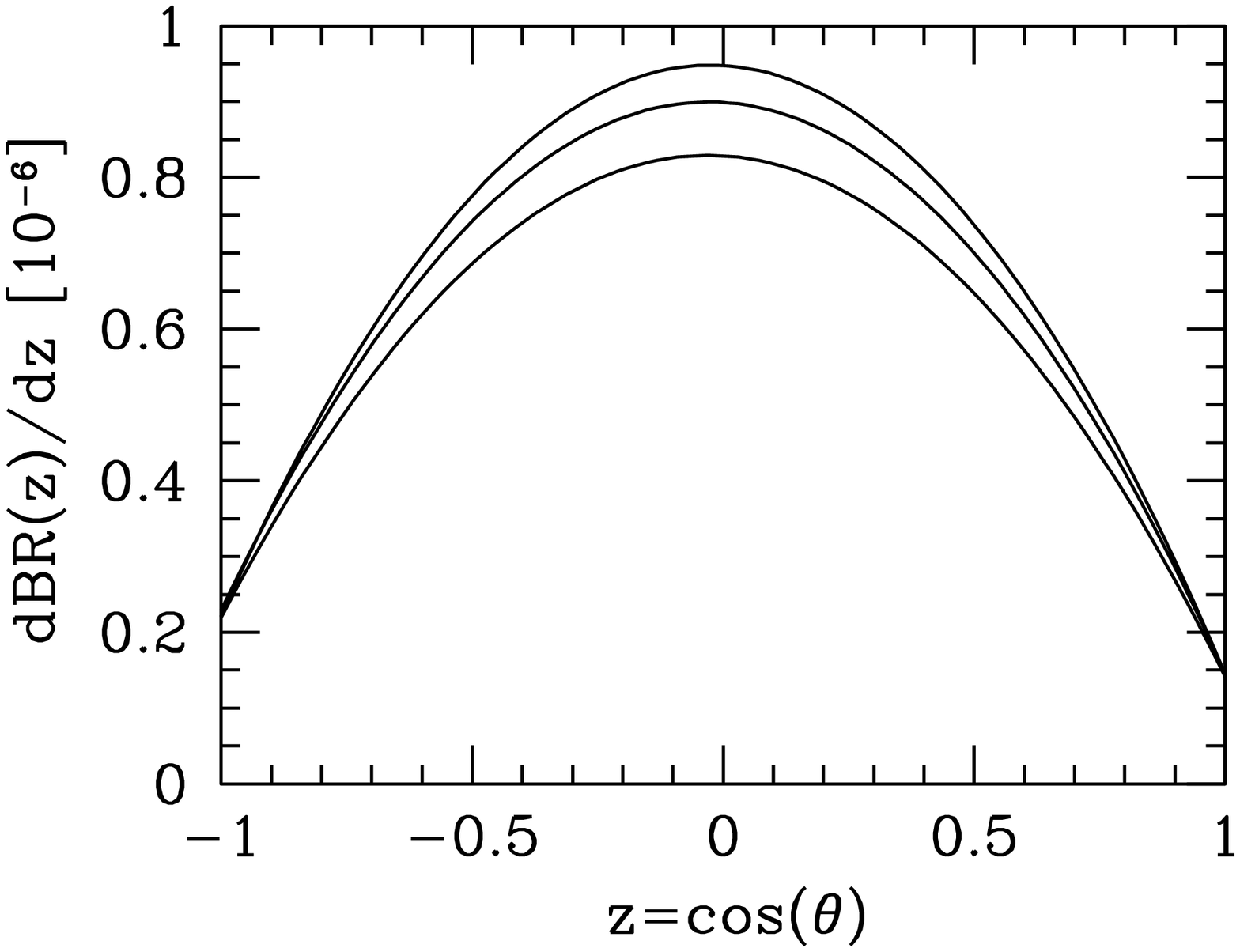}
    \vspace{0.5cm}
    \caption[]{Left frame: 
    NNLL branching ratio differential in $z=\cos \theta$ for four
    bins in $\sh$.  
    Bin 1: $0.05 \le \sh \le 0.10$ (solid);
    bin 2: $0.10 \le \sh \le 0.15$ (dotted);
    bin 3: $0.15 \le \sh \le 0.20$ (short-dashed);
    bin 4: $0.20 \le \sh \le 0.25$ (long-dashed).
    $m_c^{\pole}/m_b^{\pole} = 0.29$ and $\mu=5$ GeV. \\
    Right frame: NNLL branching ratio differential in $z=\cos \theta$. 
    $\sh$ is integrated in the interval $0.05 \le \sh \le 0.25$.  
    The curves correspond to $\mu=2.5$ GeV (lowest), $\mu=5.0$ GeV (middle)
    and $\mu=10.0$ GeV (uppermost). $m_c^{\pole}/m_b^{\pole} = 0.29$.}
    \label{fig:4}
    \end{center}
\end{figure}
We expect that in the future also the angular distribution
in $\theta$ will become measurable. 
In the left frame in fig. \ref{fig:4} we show the branching
ratio differential in the variable $z=\cos \theta$ for four
bins in $\sh$, using $\mu=5$ GeV for the renormalization scale
and putting $m_c^{\rm{pole}}/m_b^{\rm{pole}}=0.29$. In the right
frame we show this branching ratio after integrating $\sh$ over
the interval $0.05 \le \sh \le 0.25$ for three values of the
renormalization scale and putting  $m_c^{\rm{pole}}/m_b^{\rm{pole}}=0.29$.  
\section{Summary}
\label{sec:nine}
In this paper we presented NNLL results for the double differential 
decay width $d\Gamma(b \to X_s \ell^+ \ell^-)/(d\sh \, dz)$. The variable
$z$ denotes $\cos(\theta)$,
where $\theta$ is the angle between the momenta of the $b$-quark and the
$\ell^+$, measured in the rest-frame of the lepton pair. 
To obtain these results, genuinely new calculations were necessary
for the combined virtual- and gluon bremsstrahlung corrections
associated with the operators $O_7$, $O_9$ and $O_{10}$. These corrections
are encoded in the functions 
$f_{99}(\sh,z)$,
$f_{77}(\sh,z)$,
$f_{79}(\sh,z)$,
$f_{910}(\sh)$ and
$f_{710}(\sh)$ 
in the general expression (\ref{doublewidth}) 
for the double differential decay width. To obtain a NNLL prediction
for this quantity, we combined
these new ingredients with existing results
on the NNLL Wilson coefficients and on
the virtual corrections to the matrix elements 
of the operators $O_1$, $O_2$ and $O_8$. 
As the virtual QCD corrections to the matrix elements of $O_1$ and $O_2$
are only known for values of $\sh \le 0.25$, this implies that
NNLL corrections to the double differential decay width are
available only for values of $\sqrt{s}$ below the $J/\psi$ resonance.
In this paper, we neglected certain bremsstrahlung contributions, which
in principle contribute at NNLL precision. This omission is well motivated
by the fact that the corresponding corrections have  a very small impact on 
$d\Gamma(b \to X_s \ell^+ \ell^-)/d\sh$.

{}From our results on the double differential decay width we 
derived NNLL results for the lepton forward-backward 
asymmetries, as these quantities are known to be very sensitive to new physics.
We found that the NNLL corrections drastically reduce the renormalization
scale ($\mu$) dependence of the forward-backward asymmetries. 
In particular, $\sh_0$, the position at which the
forward-backward asymmetries vanish, is essentially free of uncertainties
due to the renormalization scale at NNLL precision. At NNLL precision, we
found $\sh_0^{\rm{NNLL}}=0.162 \pm 0.005$, where the error is dominated
 the uncertainty in $m_c/m_b$. This is to be compared with the NLL result,
$\sh_0^{\rm{NLL}}=0.144 \pm 0.020$, where the error is dominated by
uncertainties due to the choice of $\mu$.

\noindent
When we were working out
the double differential decay width, a paper on the  
NNLL predictions for the forward-backward asymmetries was submitted
to the hep-archive \cite{Isidori}. As these authors used
a different regularization scheme for infrared- and collinear singularities
and another procedure for the evaluation of the phase space
integrals, the two papers provide independent calculations
of the forward-backward asymmetries. Our results are in full agreement
with those presented in their final version
\cite{Isidori}.

\noindent
\underline{{\bf Acknowlegments:}} We thank 
Haik Asatrian and Manuel Walker for useful discussions.

\newpage
% -------- appendices  ------------
\appendix
\section{$\boldsymbol{\omega_{77}(\sh)$, $\omega_{99}(\sh)$ and $\omega_{79}(\sh)}$}
In this appendix we repeat the explicit expressions for
the functions $\omega_{77}(\sh)$, $\omega_{99}(\sh)$ and $\omega_{79}(\sh)$
which contain the virtual- and bremsstrahlung corrections to the
matrix elements associated with the operators
$\wtO_7$, $\wtO_9$ and $\wtO_{10}$. For their derivation, we
refer to \cite{Asatrian:2001de,Asatrian1}. 
The functions read ($\Li_2(x) = - \int_0^x \, dt/t \, \ln(1-t)$)
%----- omega_7 -----
\bea
\omega_{77}(\s)  &=&
    - \frac{8}{3}\,\ln \left( \frac{\mu}{m_b} \right)
    - \frac{4}{3}\, \Li_2(\s)
    - \frac{2}{9} \,{\pi }^{2}
    - \frac{2}{3}\, \ln(\s) \ln(1-\s) \nonumber \\
    &&
    - \frac{1}{3}\, \frac{8 + \s}{2 + \s} \ln (1-\s)
    - \frac{2}{3}\, \frac{\s \left(2 - 2\,\s - \s^2 \right)} {\left(1 - 
      \s \right)^{2} \left(2 + \s \right)} \ln(\s)
    - \frac{1}{18}\, \frac {16 - 11\,\s - 17\,\s^{2}} {\left( 2 +
      \s \right) \left( 1 - \s \right)} \, ,
\eea
%----- omega_9 -----
\bea
\label{omega99}
\omega_{99}(\hat{s}) &=&
    - \frac{4}{3}\, \Li_2(\s)
    - \frac{2}{3} \ln(1 - \s) \ln(\s)
    - \frac{2}{9}\pi^2
    - \frac{5 + 4\,\s}{3(1 + 2\,\s)} \ln(1-\s) \nonumber \\
    &&
    - \frac{2\,\s\,(1 + \s)(1 - 2\,\s)}{3\,(1 - \s)^2 (1 + 2\,\s)} \ln(\s)
    + \frac{5 + 9\,\s - 6\,\s^2}{6\,(1 - \s)(1 + 2\,\s)} \, ,
\eea
%----- omega_79 -----
\bea
\omega_{79}(\s)  &=&
    - \frac{4}{3}\,\ln \left (\frac {\mu}{m_b}\right)
    - \frac{4}{3}\, \Li_2(\s)
    - \frac{2}{9}\,{\pi }^{2}
    - \frac{2}{3}\,\ln(\s) \ln(1 - \s) \nonumber \\
    &&
    - \frac{1}{9}\,\frac{2 + 7\,\s}{\s} \ln (1-\s)
    - \frac{2}{9}\, \frac{\s \left(3 - 2\,\s \right)}{\left( 1 - 
       \s \right)^{2}} \ln(\s)
    + \frac{1}{18}\, \frac{5 - 9\,\s}{1 - \s} \, .
\eea
\section{Auxiliary quantities $\boldsymbol{A_i}$, $\boldsymbol{T_9}$, 
$\boldsymbol{U_9}$ and $\boldsymbol{W_9}$}
The auxiliary quantities $A_i$, $T_9$, $U_9$ and $W_9$ appearing in the
effective Wilson coefficients in 
eqs.~(\ref{effcoeff7})--(\ref{effcoeff10}) are the
following linear combinations of the Wilson coefficients $C_i(\mu)$ 
\cite{Bobeth:2000mk,ALGH}:
    \begin{align}
        \label{ATUW}
        A_7 =&\, \frac{4\, \pi}{\alpha_s(\mu)} \, C_7(\mu) - 
\frac{1}{3} \, C_3(\mu) - \frac{4}{9} \, C_4(\mu) -
                \frac{20}{3} \, C_5(\mu) - \frac{80}{9} \, C_6(\mu) \, , 
 \nonumber \\
        A_8 =&\, \frac{4\, \pi}{\alpha_s(\mu)} \, C_8(\mu) +  C_3(\mu) - 
\frac{1}{6} \, C_4(\mu) + 20 \, C_5(\mu) -
                \frac{10}{3} \, C_6(\mu) \, ,  \nonumber \\
        A_9 =&\, \frac{4 \pi}{\alpha_s(\mu)} \, C_9(\mu) + \sum_{i=1}^{6} \, 
C_i(\mu) \, \gamma_{i9}^{(0)} \,
                \ln \! \left( \frac{m_b}{\mu} \right)
                 + \frac{4}{3} \, C_3(\mu) + \frac{64}{9} \, C_5(\mu) + 
\frac{64}{27} \, C_6(\mu) \, ,  \nonumber \\
        A_{10} =&\, \frac{4 \pi}{\alpha_s(\mu)} \, C_{10}(\mu) \, , \\
        T_9 =&\, \frac{4}{3} \, C_1(\mu) +  C_2(\mu) + 6 \, C_3(\mu) + 60 \, C_5(\mu) \, ,
                \nonumber \\
        U_9 =& - \frac{7}{2} \, C_3(\mu) - \frac{2}{3} \,C_4(\mu) -38
        \,C_5(\mu) 
- \frac{32}{3} \,C_6(\mu) \, , \nonumber \\
        W_9 =& - \frac{1}{2} \, C_3(\mu) - \frac{2}{3} \,C_4(\mu) -8
        \,C_5(\mu) 
- \frac{32}{3} \,C_6(\mu) \, .\nonumber
    \end{align}
The entries $\gamma_{i9}^{(0)}$ of the anomalous dimension matrix
read for $i=1,...,6$: $(-32/27,-8/9,-16/9,32/27,-112/9,512/27)$.
In the contributions which explicitly involve virtual or bremsstrahlung
correction 
only the leading order coefficients
$A_i^{(0)}$, $T_9^{(0)}$, $U_9^{(0)}$ and $W_9^{(0)}$ enter. They are given by
    \begin{align}
        \label{ATUW0}
        A_7^{(0)} =&\, \, C_7^{(1)} - \frac{1}{3} \, C_3^{(0)} - 
\frac{4}{9} \, C_4^{(0)} -
                \frac{20}{3} \, C_5^{(0)} - \frac{80}{9} \, C_6^{(0)} \, ,  
\nonumber \\
        A_8^{(0)} =&\, \, C_8^{(1)} +  C_3^{(0)} - \frac{1}{6} \, 
C_4^{(0)} + 20 \, C_5^{(0)} -
                \frac{10}{3} \, C_6^{(0)} \, ,  \nonumber \\
        A_9^{(0)} =&\, \frac{4\, \pi}{\alpha_s} \left( C_9^{(0)} + 
\frac{\alpha_s}{4\,\pi}\, C_9^{(1)} \right) +
                \sum_{i=1}^{6} \, C_i^{(0)} \, \gamma_{i9}^{(0)} \,
                \ln \! \left( \frac{m_b}{\mu} \right) + \frac{4}{3} \, 
C_3^{(0)} + \frac{64}{9} \, C_5^{(0)} +
                \frac{64}{27} \, C_6^{(0)} \, ,  \nonumber \\
        A_{10}^{(0)} =&\, C_{10}^{(1)} \, , \vspace*{0.3cm} \\ 
\vspace*{0.3cm}
        T_9^{(0)} =&\, \frac{4}{3} \, C_1^{(0)} +  C_2^{(0)} + 
6 \, C_3^{(0)} + 60 \, C_5^{(0)} \, ,
                \nonumber \\
        U_9^{(0)} =& - \frac{7}{2} \, C_3^{(0)} - \frac{2}{3} \,
C_4^{(0)} -38 \,C_5^{(0)}
                - \frac{32}{3} \,C_6^{(0)} \, , \nonumber \\
        W_9^{(0)} =& - \frac{1}{2} \, C_3^{(0)} - \frac{2}{3} \,
C_4^{(0)} -8 \,C_5^{(0)}
                - \frac{32}{3} \,C_6^{(0)} \, .\nonumber
    \end{align}
We list the leading and next-to-leading order contributions to the 
quantities $A_i$, $T_9$, $U_9$ and $W_9$ in Tab.
\ref{table1}.
\def\lb{\raisebox{0.5mm}{\big(}}
\def\rb{\raisebox{0.5mm}{\big)}}
\begin{table}[hbt]
    \begin{center}
    \begin{tabular}{lccc}
        \hline\hline
        $\mu$                                & $ 2.5$ GeV          & $ 5$ GeV           & $ 10$ GeV           \\
        \hline
        $\alpha_s                          $ & $ 0.267           $ & $ 0.215          $ & $ 0.180           $ \\
        $C_1^{(0)}                         $ & $ -0.697          $ & $ -0.487         $ & $-0.326           $ \\
        $C_2^{(0)}                         $ & $ 1.046           $ & $ 1.024          $ & $ 1.011           $ \\
        $\lb A_7^{(0)},~A_7^{(1)}\rb       $ & $ (-0.360,~0.031) $ & $(-0.321,~0.019) $ & $ (-0.287,~0.008) $ \\
        $A_8^{(0)}                         $ & $ -0.164          $ & $ -0.148         $ & $ -0.134          $ \\
        $\lb A_9^{(0)},~A_9^{(1)}\rb       $ & $ (4.241,~-0.170) $ & $(4.129,~0.013)  $ & $ (4.131,~0.155)  $ \\
        $\lb T_9^{(0)},~T_9^{(1)}\rb       $ & $ (0.115,~0.278)  $ & $(0.374,~0.251)  $ & $ (0.576,~0.231)  $ \\
        $\lb U_9^{(0)},~U_9^{(1)}\rb       $ & $ (0.045,~0.023)  $ & $(0.032,~0.016)  $ & $ (0.022,~0.011)  $ \\
        $\lb W_{9}^{(0)},~W_{9}^{(1)}\rb   $ & $ (0.044,~0.016)  $ & $(0.032,~0.012)  $ & $ (0.022,~0.009)  $ \\
        $\lb A_{10}^{(0)},~A_{10}^{(1)}\rb $ & $ (-4.372,~0.135) $ & $(-4.372,~0.135) $ & $ (-4.372,~0.135) $ \\
        \hline\hline
\end{tabular}
\end{center}
\caption{\label{table1} Coefficients appearing
        in eqs.~(\ref{effcoeff7})--(\ref{effcoeff10})
for $\mu=2.5$~GeV,
$\mu=5$~GeV and $\mu=10$~GeV. For $\alpha_s(\mu)$ (in the
$\overline{\mbox{MS}}$ scheme) 
we used the two-loop
expression with five flavors and $\alpha_s(m_Z)=0.119$. The entries correspond
to the 
pole top quark mass
$m_t=174$~GeV. The superscript (0) refers to lowest order quantities while
the superscript (1) denotes the correction terms of order $\alpha_s$,
i.e. $X=X^{(0)} + X^{(1)}$ with $X=C,A,T,U,W$.}
\end{table}

Finally, we give the function $h(z,\sh)$ which appears
in the effective Wilson coefficients in 
eqs.~(\ref{effcoeff7})--(\ref{effcoeff10}):
\bea
\label{hfun}
    h(z,\s) &=& - \frac{4}{9} \ln(z) + \frac{8}{27} + 
     \frac{16}{9}\frac{z}{\s} \nonumber \\
  &&
              - \frac{2}{9} \left( 2+\frac{4\,z}{\s} \right)
              \sqrt{\left|\frac{4\,z-\s}{\s}\right|} \cdot
              \begin{cases}
                  2 \arctan \sqrt{\frac{\s}{4\,z-\s}} \,\, , & \s < 4\,z \\ \\
                  \ln \left(\frac{\sqrt{\s} + \sqrt{\s - 4\,z}}{\sqrt{\s} 
                  - \sqrt{\s - 4\,z}} \right) -i\,\pi, & \s > 4\,z
              \end{cases} \,\, .
\eea

\end{document}